\documentclass[11pt,a4paper]{article}

\usepackage{graphicx}
\usepackage{tabularx}
\usepackage{array}
\usepackage{adjustbox}
\usepackage[utf8]{inputenc}
\usepackage{booktabs}
\usepackage[hang,flushmargin]{footmisc}
\usepackage{threeparttable}
\usepackage{multirow}
\usepackage{rotating}
\usepackage[natbibapa, bibnewpage, nodoi]{apacite}
\usepackage{soul}
\usepackage{url}
\usepackage[colorlinks=false]{hyperref}
\usepackage{amsmath, amssymb}
\usepackage{tikz}
\usepackage{pgfplots}
\pgfplotsset{width=10cm,compat=1.9}
\usepackage{setspace}
\usepackage{color}
\usepackage{caption}
\usepackage{ragged2e}
\usepackage{titlesec}
\usepackage{float}
\usepackage{subcaption}
\usepackage{makecell}
\usepackage{enumitem}
\usepackage[american]{babel}
\usepackage[toc,page]{appendix}
\usepackage{chngcntr}
\usepackage{float}
\usepackage{longtable}
\usepackage{booktabs}
\usepackage{etoolbox}
\apptocmd{\sloppy}{\hbadness 10000\relax}{}{} 
\usepackage{pdflscape}
\usepackage{caption}
\usepackage{array}    
\usepackage[section]{placeins}
\usepackage{afterpage}

\usepackage[margin=1in]{geometry}
\usepackage{titling} 
\setlist[itemize]{noitemsep, topsep=0pt, left=1em}
\titleformat{\section}
  {\normalfont\Large\bfseries}
  {\thesection}
  {1em}
  {}
\titleformat{\subsection}
  {\normalfont\large\bfseries}
  {\thesubsection}
  {1em}
  {}

\renewcommand{\thefootnote}{\fnsymbol{footnote}}

\title{Emotional driving:\\
Reference-dependent emotions and risky driving behavior\\
after sporting events%
\protect\footnotemark[1]
}
\author{
\begin{tabular}{ccc}
\small Travis Richardson\protect\footnotemark[2] &
\small Steve Bickley\protect\footnotemark[3] &
\small Ho Fai (Ben) Chan\protect\footnotemark[3] \\
\small Benno Torgler\protect\footnotemark[3] &
\small Shamsunnahar Yasmin\protect\footnotemark[5] &
\small Tim Pawlowski\protect\footnotemark[4] \\
\end{tabular}
}

\vspace{2.5em}
\date{\today}

\begin{document}

\maketitle
\vspace{-2em}
\setlength{\parindent}{2em}

\begingroup
\footnotesize
\footins
\setstretch{1}
\footnotetext[2]{Corresponding author. University of Tübingen, Institute of Sports Science. Email: travis-william.richardson@uni-tuebingen.de}
\footnotetext[3]{Queensland University of Technology, School of Economics \& Finance. Emails: steve.bickley@qut.edu.au, hofia.chan@qut.edu.au, benno.torgler@qut.edu.au}
\footnotetext[4]{University of Tübingen, Institute of Sports Science, LEAD Graduate School and Research Network, Interfaculty Research Institute for Sports and Physical Activity. Email: tim.pawlowski@uni-tuebingen.de}
\footnotetext[5]{Queensland University of Technology, Centre for Accident Research and Road Safety. Email: shams.yasmin@qut.edu.au}
\footnotetext[1]{Data was kindly provided by HERE - Road Traffic Analytics Location Intelligence. HERE had no role in the study design, further data collection and analysis, or the preparation of the manuscript. The authors do not have a financial interest in the topic of this paper and there are no conflicts of interest. The authors made use of AI-assisted tools (Claude, Anthropic) for coding assistance in data processing and regression analysis, as well as for language and text editing support during manuscript preparation. All substantive intellectual contributions, analytical decisions, and conclusions remain those of the authors. All errors are our own.}
\endgroup

\vspace{2.5em}
\vspace{-1em}
\begin{abstract}
\footnotesize
\setstretch{1}
\noindent
Using average vehicle speed data in 10-minute increments at the Traffic Message Channel (TMC) location level, along with precise crash timing and location information, we analyze driving behavior around five Florida stadiums before and after NFL and NBA regular season games from 2015 to 2019. We find no evidence of emotional driving following NBA games, but strong and consistent effects following NFL games, concentrated in predicted-close games that end in disappointing home-team losses --- combining high pre-game suspense with negative outcome valence. These games are associated with significant increases in average vehicle speed within 3 km of stadiums during the first post-game hour, dissipating with increasing time and distance from the stadium. Average vehicle speed increases by up to 3 mph relative to predicted-close games that ended in a win --- an effect several times larger than the typical game day versus non-game day speed differential. Overall, our results highlight how the combination of sustained suspense and negative outcome valence in close sporting contests can spill over into risky post-game driving behavior, underscoring the behavioral and public safety implications of affective cues in large-scale sporting events.

\vspace{2.5em}
\end{abstract}

\clearpage
\pagenumbering{arabic}
\renewcommand{\thefootnote}{\arabic{footnote}}
\setcounter{page}{1}
\setstretch{1.5}
\section{Introduction}

A central insight from models of reference-dependent preferences is that outcomes are evaluated not only in absolute terms, but relative to expectations \citep{Kahneman1979}. When realized outcomes deviate from an expectations-based reference point, individuals may experience elation when outcomes exceed expectations, or disappointment, anger, and frustration when outcomes fall short \citep{Bell1985, Loomes1986, vanDijk1997}, and these affective responses may shape subsequent behavior.

According to the gain-loss utility framework by \citet{Koszegi2006}, such deviations from the expectations-based reference point might generate rapid and strong behavioral responses. Moreover, emotions generated by economic and social experiences can spill over into decisions that are seemingly unrelated to the original source of affect. Sporting events provide a uniquely clean natural setting for observing such behavior in the field: fans enter games with well-defined expectations and hopes about the likely or preferred outcome, and their emotional reactions to deviations from that set point, whether positive or negative, can spill over into behaviors beyond the game itself. Recent evidence has documented such emotional spillovers in a variety of contexts, including stock market returns \citep{Edmans2007}, voter turnout \citep{Potoski2017}, judicial sentencing \citep{Eren2018}, consumer spending \citep{CaiLi2026}, and intimate partner violence \citep{Card2011, Cardazzi2024}.

This paper investigates whether and how emotionally charged game outcomes might affect post-game driving behavior using fine-grained spatiotemporal data on vehicle speed in Florida. Driving behavior is a particularly valuable outcome for studying emotional spillovers for several reasons. First, post-game travel by car is the dominant mode of transportation to and from stadium events, meaning a large share of emotionally affected sports fans transition directly from the stadium into real-world traffic \citep{Humphreys2018}. Second, unlike outcomes such as domestic violence or consumer spending, vehicle speed is observable at high temporal and spatial resolution, allowing for measurable minute-by-minute and kilometer-by-kilometer (mile-by-mile) emotional responses as fans disperse from the venue \citep{Hwang2025}. Third, even small deviations in vehicle speed carry meaningful public safety consequences; according to the \citet{WHO2023}, each 1\% increase in average speed results in a 4\% increase in the likelihood of a fatal crash, making driving a particularly high-stakes domain in which to study emotional spillovers.

The growing field of affective automotive user interfaces recognizes that a driver's emotional state profoundly influences road safety \citep{Braun2021}. Literature extensively reviewed by \citet{Braun2021} highlights how medium arousal and positive valence typically support optimal driving. This is in line with the Yerkes-Dodson law \citep{Yerkes1908}, which holds that performance peaks at moderate levels of arousal, but declines when arousal becomes too low or too high. Negative emotions have clear adverse effects, but even positive emotions can impair performance under high arousal \citep{Cai2011, Russell1980}, as elevated arousal can disrupt attentional focus and impair the cognitive control mechanisms that support safe driving.\footnote{Anger is consistently linked to aggressive and reckless driving and worsens performance by increasing reaction times and impairing risk perception \citep{Mesken2007, Underwood1999, Precht2017, Zhang2016, Zimasa2019, Deffenbacher2003, Jeon2014, Roidl2014, Steinhauser2018}. Even high-arousal happiness can increase risky behavior \citep{TaubmanBenAri2012}. \citet{Zhang2020} find that both angry and happy drivers leave less space between vehicles and brake more slowly, while \citet{Zhang2022} show that anger increases speed and risk near pedestrians.} In a naturalistic study of taxi drivers in Japan, \citet{Kadoya2021} find that negative emotional states, anger and sadness, significantly increase driving speed in real traffic conditions, while happy and relaxed states show no such effect. Crucially, \citet{Dingus2016} show that emotional drivers, whether angry or sad, are nearly 10 times more likely to crash compared to baseline driving conditions. In summary, driving appears particularly sensitive to emotional states because emotional arousal can impair risk perception and increase risky behavior. However, large-scale field evidence about the effects of emotional cues, like emotionally salient game outcomes, on driving patterns is largely missing. 

One prime exception is the study by \citet{Wood2011}, who analyze major college and professional football and basketball games and show that close, competitive contests are associated with significantly higher traffic fatalities near the winning team's location. They interpret this as evidence that physiological arousal, potentially driven by the excitement and hormonal effects of a narrow win, translates into aggressive driving behavior. Their findings emphasize the importance of emotional intensity, not just outcome valence, in shaping traffic risk. Crucially, however, Wood et al. focus on winning fans and celebratory arousal. Whether the frustration and disappointment experienced by fans of the losing team generate similarly elevated driving risk remains an open question; one that aggregate, event-level crash data are ill-suited to answer. Building on this insight, our study focuses on the behavioral mechanism underlying such effects by examining how emotional cues from sporting outcomes influence post-game vehicle speed in real-world traffic. Because emotional arousal is transient by nature, peaking immediately after the game and dissipating over time and distance, identifying its behavioral imprint requires data that are disaggregated both temporally and spatially. Using fine-grained speed data measured at 10-minute intervals across distance bands of 1 km, we provide direct evidence of how emotional arousal after emotionally salient contests manifests in driving behavior, and how quickly and how far those effects travel from the stadium.\footnote{\citet{Redelmeier2003} observe a 41\% spike in traffic fatalities after the Super Bowl, while \citet{Jakar2023} report a significant increase in vehicular crashes with reported damage in Cleveland, Ohio, particularly after highly attended NFL games. In contrast to these event-level studies focused on aggregate crash outcomes, our primary focus is put on minute-level vehicle speed as a behavioral mechanism that underlies elevated post-game risk. Our main analysis, however, will be complemented with a crash analysis (see Section~\ref{sec:crash}).}

We test what we refer to as \textit{emotional driving}, i.e., observable changes in driving behavior that occur when individuals transition from emotionally intense settings, such as witnessing unexpected or unwanted game outcomes, into real-world traffic environments. Close, tightly contested games are especially relevant, as they heighten emotional arousal and amplify reactions to final outcomes. Specifically, we explore whether and how predicted close games that ended in a loss, as well as upset wins and losses by the home team of professional National Basketball Association (NBA) and National Football League (NFL) games, might lead to measurable changes in post-game driving speed of vehicles in close geographic proximity to the stadiums where the matches are played. 

For our analysis, we link Global Positioning System (GPS)-based traffic data surrounding five stadiums and arenas in Florida with information about the corresponding NFL and NBA home teams as well as their regular season games---including pre-game betting spreads to construct ex-ante expectations about the likely game outcome---from 2015 to 2019, a period unaffected by the COVID-19 pandemic.\footnote{We focus on the 2015--2019 period to ensure a consistent, pre-pandemic sample. The COVID-19 pandemic beginning in 2020 fundamentally disrupted stadium attendance, fan behavior, and traffic patterns, making post-2019 seasons ill-suited for comparison.} Our fixed-effects regression models are disaggregated by time (60/120 minutes before/after the games, 10-minute increments) and distance (0--5 kilometers around the stadium, 1-km increment).\footnote{Technically, only the lower bound of each distance band is an integer; for example, a radius labeled ``1--2 km'' corresponds to distances from 1.0 km up to but not including 2.0 km. For ease of reading, we retain integer values for both bounds throughout the text, tables, and figures.}

Overall, our analysis reveals no evidence of emotional driving following NBA games. This could be attributed to comparably small attendance numbers or arenas being located downtown in metropolitan areas with dense traffic (hence, there could be less opportunity/environmental affordance to speed). Another plausible explanation could be the high-scoring nature of basketball making any upsets less likely and outcome-related emotions less intense.\footnote{We further elaborate on these possible explanations in the discussion section.} For the NFL, however, we find that the strongest and most consistent effects occur after predicted close games that end in disappointing home-team losses. These contests, which combine high suspense with emotional frustration, are associated with significant increases in average vehicle speed. These effects are concentrated in the first hour after the game and decline with increasing time and distance to the stadium. This pattern is robust across different specifications, alternative spread thresholds, sample restrictions, controls for belief updating, and is not replicated when game outcome labels are randomly reassigned or when real game labels are applied to non-game Sundays.

This paper contributes to three strands of literature. First, it contributes to behavioral economics by providing field evidence that expectations-based emotional cues can spill over into immediate real-world behavior, complementing laboratory and survey evidence on reference-dependent preferences \citep{Koszegi2006, Bell1985, Loomes1986}. Second, it contributes to research on sports outcomes and social behavior by identifying a high-frequency behavioral mechanism, minute-level vehicle speed, that may help explain why emotionally intense games are associated with elevated public safety risks \citep{Wood2011, Card2011, Cardazzi2024}. Third, it contributes to research on transportation and road safety by showing that risky driving behavior can be shaped not only by road conditions and enforcement, but also by predictable affective states generated by collective events \citep{Dingus2016, Kadoya2021}.

Taken together, these findings emphasize that high-arousal sporting results in the NFL, specifically close, disappointing contests, can translate into tangible risks on the road which should be considered for traffic management and public safety near stadiums. Simple policy interventions, such as targeted speed enforcement, improved signage, increased public transportation, or post-game crowd management, might reduce the likelihood of emotionally driven vehicle speed increases in these high-risk windows.

The remainder of this paper is structured as follows: Section~\ref{sec:framework} develops a conceptual framework distinguishing surprise, suspense, valence, and behavioral response; Section~\ref{sec:data} describes the data---including traffic, weather, and sports game information---and outlines the empirical strategy; Section~\ref{sec:results} presents the main findings and several robustness checks; Section~\ref{sec:discussion} offers a discussion of the limitations and implications of our study; Section~\ref{sec:conclusion} concludes.

\section{Conceptual framework}
\label{sec:framework}

Our analysis is motivated by the distinction between surprise, suspense, and outcome valence. Upsets represent deviations from pre-game expectations: the home team surprisingly wins when a loss was expected, or loses when a win was expected. Close games, by contrast, may generate high suspense even when the final outcome is not statistically surprising, since either team had a plausible chance of winning before the game. In our empirical setting, betting-market point spreads provide a useful proxy for pre-game expectations, with games classified as predicted wins, predicted close contests, or predicted losses from the perspective of the home team (see Section~\ref{sec:sportsdata} for formal definitions).

To clarify the behavioral mechanism, we distinguish three related but conceptually separate emotional dimensions --- surprise, suspense, and valence --- and one behavioral dimension: behavioral response. All three emotional dimensions are grounded in pre-game expectations, which form the reference point against which outcomes are evaluated \citep{Koszegi2006, Koszegi2009}. When outcomes fall short of expectations, fans may experience disappointment, frustration, or anger; when outcomes exceed expectations, fans may experience elation or relief \citep{Bell1985, Loomes1986}.

\textit{Surprise} refers to the deviation between the expected and the realized outcome: a game ends in a surprise when the home team wins despite being predicted to lose, or loses despite being predicted to win \citep{Koszegi2006}. An upset loss combines negative valence with surprise; an upset win combines positive valence with surprise. 

Surprise is not the same as suspense, however, a distinction between the two is important to note \citep{Ely2015}: a game can be highly surprising without having been suspenseful (e.g., a heavily favored team losing a lopsided contest), or highly suspenseful without being surprising (e.g., a predicted-close game decided narrowly). A predicted-close loss may therefore not be highly surprising in a probabilistic sense, yet it combines sustained emotional arousal throughout the contest with a negative final resolution --- so the behavioral response may be driven not only by the deviation from expectations, but also by the intensity of suspense experienced during the game.

\textit{Suspense} refers to the emotional intensity generated while the outcome remains unresolved \citep{Ely2015}. A game expected to be close may sustain attention and arousal throughout the contest, since fans remain uncertain about the final result until late in the game. In our empirical setting, we proxy for sustained suspense using the pre-game point spread, under the assumption that games predicted to be close are more likely to remain competitive throughout. We partially test this assumption in Section~\ref{sec:halftimeupdate}, where we examine whether halftime score differentials alter the main findings.

\textit{Valence} refers to whether the final outcome is emotionally positive or negative for the home-team fan base \citep{Russell1980}: a home-team win is likely to generate happiness, while a home-team loss is likely to generate disappointment, frustration, or anger.

\textit{Behavioral response} refers to the observable post-game driving behavior that may follow from these emotional states. The relationship between valence and behavioral response is not simply positive versus negative, however. In a judgment and decision-making context, \citet{Lerner2001} show that anger and fear, though both negative emotions, have opposing effects on risk-taking, with angry individuals' risk assessments more closely resembling those of happy individuals than fearful individuals. In a simulated driving context, however, \citet{Jeon2014} find that while both anger and happiness degrade driving performance overall, only anger significantly affects aggressive driving, including speed, infractions, and collisions, whereas happiness does not. Because anger intensifies with the degree of frustration experienced, a home-team loss, especially a close loss, which combines frustration and negative arousal, may therefore produce more aggressive behavioral responses than a lopsided loss, which generates less sustained arousal prior to the outcome. Emotional arousal can affect attention, risk perception, self-control, and aggressiveness behind the wheel \citep{Dingus2016, Kadoya2021}. In our setting, these responses are measured through changes in average vehicle speed near stadiums after games. Because the traffic data are aggregated, we do not observe individual fans' emotions or driving choices directly. The analysis therefore estimates reduced-form changes in local driving behavior around emotionally salient events.

Table~\ref{tab:framework} summarizes the emotional profiles of each outcome type, combining pre-game expectations, actual results, and the resulting surprise--suspense--valence profiles.

\begin{table}[H]
\centering
\refstepcounter{table}
\label{tab:framework}
\captionsetup{justification=centering, font=small, labelfont=bf, labelsep=colon}
\caption*{\textbf{Table \thetable}\\[4pt]Emotional profiles of game outcome types}
\vspace{6pt}
\scriptsize
\setlength{\tabcolsep}{4pt}
\renewcommand{\arraystretch}{.7}
\begin{tabular*}{\textwidth}{@{\extracolsep{\fill}}llll@{}}
\toprule
\multicolumn{1}{l}{\textbf{Expectation}} & \multicolumn{1}{c}{\textbf{Outcome}} & \multicolumn{1}{l}{\textbf{Emotional profile}} & \multicolumn{1}{l}{\textbf{Interpretation}} \\
\midrule
Predicted win   & Home loss & Surprise and negative valence & Upset loss; disappointment \\
Predicted close & Home loss & Suspense and negative valence & High-suspense, negative resolution \\
Predicted loss  & Home win  & Surprise and positive valence & Upset win; elation \\
Predicted close & Home win  & Suspense and positive valence & High-suspense, positive resolution \\
Predicted win   & Home win  & Expected positive outcome    & Confirmation of favorable expectations \\
Predicted loss  & Home loss & Expected negative outcome    & Confirmation of unfavorable expectations \\
\bottomrule
\end{tabular*}
\noindent
\begin{minipage}{\textwidth}
  \vspace{10pt}
  \begin{spacing}{0.9}
  \scriptsize
  \noindent\textit{Notes:} The table classifies each game outcome type by its pre-game expectation category, actual result, and the resulting emotional profile experienced by home-team fans. Predicted win, predicted close, and predicted loss are defined using pre-game betting-market point spreads as described in Section~\ref{sec:sportsdata}. Suspense refers to sustained uncertainty during the contest; surprise refers to the deviation of the realized outcome from the pre-game reference point; valence refers to whether the final outcome is emotionally positive or negative from the perspective of the home team.
  \end{spacing}
\end{minipage}
\end{table}

While both predicted-close wins and losses involve the same level of pre-game suspense, they differ fundamentally in how that suspense resolves. A predicted-close win resolves with positive affect --- relief or celebration --- whereas a predicted-close loss resolves with frustration, disappointment, and anger, the emotions most strongly linked to aggressive driving above. The behavioral response may therefore be strongest when sustained suspense is resolved negatively.

\section{Data and methods}
\label{sec:data}

\subsection{Vehicle and weather data}
\label{sec:vehicledata}

Our primary dataset for vehicle speed, crash, and location was obtained from HERE Technologies.\footnote{HERE Technologies is a location data and technology company, best known in the automotive sector. It helps users and automated driving applications navigate with complete, accurate maps of real-world places.} HERE Probe Data is a location-based dataset that captures the movement of vehicles through anonymized and aggregated GPS signals collected from connected vehicles and mobile devices. The data provides a detailed, real-time and historical view of traffic patterns and vehicular mobility. By leveraging a large volume of spatial and temporal observations, HERE Probe Data enables analysis of congestion dynamics, travel speeds, and roadway usage. It is commonly used in applications such as traffic flow modeling, safety assessments, transportation infrastructure planning, and the enhancement of driver assistance systems. Using Global Navigation Satellite System (GNSS) or GPS, the position of each vehicle is established using satellite-based positioning reliable enough to indicate precisely which road a vehicle is on.

For our analysis, we use historical traffic flow (HERE Traffic API, Flow Resource) and incident (HERE Traffic API, Incidents Resource) data aggregated to Traffic Message Channel (TMC) level by the day of week and the time of day (10-minute increments) for all road links/corridors within a 5-kilometer radius around Florida's professional football stadiums and basketball arenas for the period 2015 to 2019.\footnote{TMC coverage is limited to major roads: ``TMCs are generally longer [comprising multiple links] and available only for important roads'' \citep{HERE2024}. Smaller residential streets and low-volume roads are therefore typically not represented in the data, as they do not generate enough continuous probe data to produce reliable speed profiles.} Each observation in our dataset corresponds to a unique TMC road segment and 10-minute time interval. These facilities include the Raymond James Stadium (Tampa Bay Buccaneers), the Hard Rock Stadium (Miami Dolphins), the EverBank Field (Jacksonville Jaguars), the American Airlines Arena (Miami Heat), and the Kia Center (Orlando Magic).\footnote{To visualize the corresponding road network coverage, ArcGIS-based maps with concentric 1 km rings (from 0--1 km through 4--5 km) centered around the arena and stadium location are presented in Appendix Figures~\ref{fig:map_bucs}--\ref{fig:map_magic}.}

Vehicle speed data originally consisted of billions of observations across the five selected stadium regions. To focus on periods most relevant to emotional responses following professional sports events, we restricted the dataset to specific temporal windows aligned with game schedules. For NFL games, we include Sundays during the 2015--2019 regular seasons, limited to one hour before the earliest kickoff and up to two hours after the latest game ending, while excluding late-night games as differing travel conditions, lighting, and crowd demographics at those times introduce confounds unrelated to emotional state. For NBA games, we focus on the 2015/16 to 2018/19 regular seasons, including games with start times of 6:00 PM or later, and retained observations account for pre- and post-game periods of 1 hour pre-game and 2 hours post-game.We adopt a one-hour pre-game and two-hour post-game window to capture behavioral buildup and aftermath effects while minimizing unrelated temporal noise. The pre-game hour reflects the typical onset of fan and media activity, whereas the two-hour post-game window accounts for extended audience or mobility responses following game completion. 

While our analysis is not primarily focused on vehicular crashes, it is important to consider crashes for isolating emotionally driven behavioral responses from physical traffic disruptions. Moreover, exploring the probability of crashes on gamedays may further add some credibility to the findings we observe in our speed analysis. We obtained statewide crash data for Florida covering the full period from January 1, 2015 to December 31, 2019. These records were filtered to match the temporal scope of the speed data and were spatially merged to corresponding TMC location segments using precise latitude and longitude coordinates, along with exact crash timestamps, enabling integration at the level of individual road segments and time intervals.

Finally, weather data were obtained from WorldWeatherData.com at 15-minute intervals for the geographic areas surrounding each of the five primary stadiums analyzed in this study. These observations were matched temporally and spatially to the corresponding traffic speed data, which were recorded in 10-minute increments. To control for environmental conditions that may affect driving behavior, we constructed binary indicator variables for adverse weather conditions: hot, cold, windy, precipitation, and low visibility. Each indicator represents values approximately one standard deviation above or below the local mean (e.g., for temperature or wind speed), with the exception of precipitation, which is coded as a binary variable indicating any recorded rainfall. Visibility is measured on a 1--10 scale, with values at or below 3---representing moderate to dense fog---classified as low visibility. These weather controls are included to isolate emotional effects on driving from physical or environmental factors.

Table~\ref{tab:descriptive_stats} reports descriptive statistics for average vehicle speed by sport, distance band, and day type. Baseline speeds range from approximately 23--29 mph for the NFL and 21--28 mph for the NBA depending on distance band, reflecting the mix of road types and traffic conditions surrounding each venue. On average, game day speeds are slightly lower than non-game day speeds across most distance bands, consistent with standard pre- and post-game congestion effects. Notably, the average difference between game day and non-game day speeds is less than 1 mph across all distance bands and both sports, providing a useful baseline against which to assess the magnitude of any emotionally induced speed changes identified in our analysis.

\begin{table}[htbp]
\centering
\refstepcounter{table}
\label{tab:descriptive_stats}
\captionsetup{justification=centering, font=small, labelfont=bf, labelsep=colon}
\caption*{\textbf{Table \thetable}\\[4pt]Descriptive statistics: average vehicle speed by sport, distance band, and day type}
\vspace{6pt}
\scriptsize
\setlength{\tabcolsep}{4pt}
\renewcommand{\arraystretch}{1}
\begin{tabular*}{\textwidth}{@{\extracolsep{\fill}}llcccccc@{}}
\toprule
& & \multicolumn{3}{c}{\textbf{Game Day}} & \multicolumn{3}{c}{\textbf{Non-Game Day}} \\
\cmidrule(lr){3-5} \cmidrule(lr){6-8}
\textbf{Sport} & \textbf{Distance band} & \textbf{Mean} & \textbf{SD} & \textbf{N} & \textbf{Mean} & \textbf{SD} & \textbf{N} \\
\midrule
NFL & 0--1 km  & 23.56 & 15.15 & 192,462   & 24.44 & 15.60 & 310,442   \\
    & 1--2 km  & 22.42 & 14.24 & 366,808   & 22.99 & 14.43 & 597,631   \\
    & 2--3 km  & 25.59 & 15.94 & 608,661   & 25.86 & 16.05 & 988,050   \\
    & 3--4 km  & 28.69 & 16.06 & 822,410   & 28.71 & 16.09 & 1,342,588 \\
    & 4--5 km  & 29.31 & 17.28 & 1,076,313 & 29.20 & 17.35 & 1,743,316 \\
\midrule
NBA & 0--1 km  & 20.64 & 16.74 & 1,019,476 & 21.06 & 16.99 & 3,557,765 \\
    & 1--2 km  & 23.00 & 14.83 & 1,580,465 & 23.19 & 15.00 & 5,522,594 \\
    & 2--3 km  & 25.47 & 13.88 & 1,380,502 & 25.64 & 14.02 & 4,828,122 \\
    & 3--4 km  & 27.65 & 14.39 & 1,438,816 & 27.85 & 14.53 & 5,024,329 \\
    & 4--5 km  & 27.97 & 12.98 & 1,201,460 & 28.11 & 13.08 & 4,212,559 \\
\bottomrule
\end{tabular*}

\noindent
\begin{minipage}{\textwidth}
  \vspace{10pt}
  \begin{spacing}{0.9}
  \scriptsize
  \noindent\textit{Notes:} Average vehicle speed is measured in miles per hour (mph). Each observation corresponds to a unique TMC road segment and 10-minute time interval. Game day observations are limited to the temporal windows aligned with game schedules as described in Section~\ref{sec:vehicledata}. Non-game day observations are matched to the same time of day and day of week as game day observations. Distance bands reflect straight-line distance from the stadium or arena. NFL statistics are based on Sunday regular-season home games for the Miami Dolphins, Tampa Bay Buccaneers, and Jacksonville Jaguars across the 2015--2019 seasons, excluding the 2018 and 2019 Miami Dolphins seasons. NBA statistics are based on regular-season home games for the Miami Heat and Orlando Magic across the 2015/16--2018/19 seasons, with tip-off times of 6:00 PM or later. The larger ratio of non-game day to game day observations for the NBA reflects the fact that NBA games occur across all days of the week, expanding the pool of matched non-game day observations relative to the NFL, where games are restricted to Sundays.
  \end{spacing}
\end{minipage}
\end{table}

\subsection{Sports data and sample selection}
\label{sec:sportsdata}

Our sample includes all regular-season home games for Florida-based teams in the NFL and the NBA during the period 2015 to 2019, comprising five NFL seasons and four NBA seasons.\footnote{The sample is restricted to regular-season games for two reasons. First, playoff games generate atypical levels of emotional intensity, fan travel patterns, and media attention that are not directly comparable to regular-season contests, potentially confounding identification. Second, across the three NFL franchises and two NBA franchises over the sample period, only two teams qualified for the playoffs, yielding just one usable NFL home playoff game and seven usable NBA home playoff games.} Specifically, we include three NFL franchises: Tampa Bay Buccaneers, Miami Dolphins, and Jacksonville Jaguars, as well as two NBA franchises: Miami Heat and Orlando Magic; yielding data for 84 NFL Sunday home games\footnote{Due to missing HERE data, we exclude the Miami Dolphins 2018 and 2019 seasons, resulting in two seasons not being included.} (with kickoff times between 1:00 PM and 4:25 PM) and 308 NBA home games (with start times of 6:00 PM or later).\footnote{Based on the \textit{NFL Follower Map}, which analyzes data from nearly 60 million Twitter followers of NFL teams and displays the most-followed team per county, we assume the local NFL team represents the dominant fan base in each region (see Appendix Figures~\ref{fig:nfl_fans} and~\ref{fig:florida_nfl_fans}; Source: \citet{SorensonNFL}). Likewise, based on the \textit{NBA Fan Map}, which analyzes data from over 70 million Twitter followers of NBA teams and displays the most-followed team per U.S. county and Canadian census division, we assume the local NBA team represents the dominant fan base in each region (see Appendix Figures~\ref{fig:nba_fans} and~\ref{fig:florida_nba_fans}; Source: \citet{SorensonNBA}).} For each game, we record the geographic coordinates of the stadium, halftime scores, final scores, and pre-game betting lines.

We closely follow \citet{Card2011} for the NFL and \citet{Cardazzi2024} for the NBA and use the point spread---widely viewed as an unbiased and informationally efficient predictor of game results \citep[see][]{Pankoff1968, Gandar1988}---as our primary measure of pre-game expectations. Accordingly, NFL games are classified based on the Las Vegas point spread as:

\begin{itemize}
\item Predicted win: Home team favored by more than 4 points (spread $\leq$ -4),
\item Predicted close: -4 $<$ point spread $<$ +4,
\item Predicted loss: Home team expected to lose by more than 4 points (spread $\geq$ +4).\footnote{\citet{Card2011} justify these cutoffs by showing that a -4 point spread corresponds to a 63\% home win probability, while a +4 point spread corresponds to 37\%, making them reasonable thresholds for differentiating expectations.}
\end{itemize}

For NBA games, we follow the classification proposed by \citet{Cardazzi2024}, who argue that the NFL's $\pm$4 threshold does not account for the higher scoring volatility and tighter margins in professional basketball. Instead, they propose using asymmetric thresholds derived from the empirical distribution of NBA point spreads:

\begin{itemize}
\item Predicted win: Home team favored by more than 6.75 points (spread $\leq$ -6.75),
\item Predicted close: -6.75 $<$ point spread $<$ +4.75,
\item Predicted loss: Home team expected to lose by more than 4.75 points (spread $\geq$ +4.75).\footnote{These cutoffs correspond to approximate home win probabilities of 70\% and 35\%, respectively, and more accurately reflect the emotional expectations embedded in basketball betting lines.}
\end{itemize}

These classifications not only define pregame expectations but also provide a natural measure of the emotional profiles described in Section~\ref{sec:framework}. In particular, predicted close games are likely to produce the strongest affective responses, as sustained uncertainty about the outcome keeps fan arousal high until the final moments, generating the high-suspense conditions that, when resolved positively or negatively, may spill over into post-game driving behavior. Predicted wins and losses that end in upsets, by contrast, combine surprise with either positive or negative valence, generating a distinct emotional profile. Tables~\ref{tab:nfl} and~\ref{tab:nba} provide a detailed description of all games for both NFL and NBA.\footnote{To facilitate scalable and reliable data management, we stored and processed all data in Apache Parquet format. Parquet is a columnar, compressed storage format which offers substantial performance advantages compared to CSV reducing the computational overhead associated with parsing, casting, and reformatting raw data, which is particularly valuable for high-frequency panel data across multiple road segments. The use of Parquet ensured not only improved processing speed during iterative filtering, merging, and regression modeling, but also preserved data integrity, particularly for date-time formatting and road segment identifiers. As a result, the Parquet format played a key infrastructural role in making our empirical pipeline reproducible, memory-efficient, and suitable for large-scale traffic behavior analysis.}

\begin{table}[htbp]
\centering
\refstepcounter{table}
\label{tab:nfl}
\captionsetup{justification=centering, font=small, labelfont=bf, labelsep=colon}
\caption*{\textbf{Table \thetable}\\[4pt]NFL games}

\vspace{6pt}

\scriptsize
\setlength{\tabcolsep}{0pt}
\renewcommand{\arraystretch}{1}
\begin{tabular*}{\textwidth}{@{\extracolsep{\fill}}lccccccc@{}}
\toprule
Team & \makecell{Total \# of \\games} & \makecell{Predicted\\win} & \makecell{Predicted\\close} & \makecell{Predicted\\loss} & \makecell{Predicted win\\$\times$ lose} & \makecell{Predicted close\\$\times$ lose} & \makecell{Predicted loss\\$\times$ win} \\
\midrule
Tampa Bay Buccaneers & 33 & 4 & 26 & 3 & 2 & 13 & 1 \\
Miami Dolphins       & 19 & 3 & 13 & 3 & 0 & 5  & 2 \\
Jacksonville Jaguars & 32 & 9 & 19 & 4 & 3 & 16 & 3 \\
\cmidrule(r){1-8}
Total NFL            & 84 & 16 & 58 & 10 & 5 & 34 & 6 \\
\bottomrule
\end{tabular*}

\noindent
\begin{minipage}{\textwidth}
  \vspace{10pt}
  \begin{spacing}{0.9}
  \scriptsize
  \noindent\textit{Notes:} Data include regular-season home games for the Miami Dolphins, Tampa Bay Buccaneers, and Jacksonville Jaguars played on Sundays with kickoff times between 1:00 PM and 4:25 PM Eastern Time. No primetime or late games are included. Seasons 2018 and 2019 for the Miami Dolphins are excluded due to missing traffic data. In addition, two games are omitted: one New Year's Eve game for the Tampa Bay Buccaneers and one for the Miami Dolphins.
  \end{spacing}
\end{minipage}
\end{table}

\begin{table}[htbp]
\centering
\refstepcounter{table}
\label{tab:nba}
\captionsetup{justification=centering, font=small, labelfont=bf, labelsep=colon}
\caption*{\textbf{Table \thetable}\\[4pt]NBA games}

\vspace{6pt}

\scriptsize
\setlength{\tabcolsep}{0pt}
\renewcommand{\arraystretch}{1}
\begin{tabular*}{\textwidth}{@{\extracolsep{\fill}}lccccccc@{}}
\toprule
Team & \makecell{Total \# of\\games} & \makecell{Predicted\\win} & \makecell{Predicted\\close} & \makecell{Predicted\\loss} & \makecell{Predicted win\\$\times$ lose} & \makecell{Predicted close\\$\times$ lose} & \makecell{Predicted loss\\$\times$ win} \\
\midrule
Miami Heat       & 153 & 42 & 101 & 10 & 13 & 43 & 3 \\
Orlando Magic    & 155 & 20 & 97  & 38 & 5  & 50 & 13 \\
\cmidrule(r){1-8}
Total NBA        & 308 & 62 & 198 & 48 & 18 & 93 & 16 \\
\bottomrule
\end{tabular*}

\noindent
\begin{minipage}{\textwidth}
  \vspace{10pt}
  \begin{spacing}{0.9}
  \scriptsize
  \noindent\textit{Notes:} Data include regular-season home games for the Miami Heat and Orlando Magic from the 2015/16 through 2018/19 NBA seasons. Games with scheduled tip-off times before 6:00 PM are excluded (17 games in total). The sample therefore consists only of games beginning between 6:00 PM and 8:00 PM Eastern Time.
  \end{spacing}
\end{minipage}
\end{table}

\subsection{Empirical strategy}
\label{sec:empirical}

Our main dependent variable is the average vehicle speed, $\text{Speed}_{it}$, measured on road segment $i$ at time $t$ (measured in 10-minute increments), within a 5-kilometer radius (stratified by distance from stadium in 1 km increments) of a home NFL or NBA stadium.\footnote{While speed is reported in miles per hour (mph), we use kilometers for spatial measurement because the smaller unit size allows for finer-grained distance bins and clearer delineation of stadium proximity effects.} To estimate how game outcomes shape driving behavior across different suspense--valence profiles, we employ a fixed-effects linear regression model that accounts for both spatial and temporal heterogeneity. Specifically, we estimate the following specification:

\begin{equation}
\text{Speed}_{it} = \alpha_i + f(p_{it}, y_{it}; \lambda) + \theta_{it} + \delta_{it} + \varepsilon_{it}
\label{eq:main}
\end{equation}

In this framework, $\alpha_i$ denotes TMC-level fixed effects to control for time-invariant characteristics of each road segment, absorbing stable differences in baseline speed across road segments such as road type, speed limit, and location. $\theta_{it}$ denotes a binary indicator for crash presence on segment $i$ at time $t$, which we include in some model specifications to distinguish behavioral changes from physical disruptions in traffic flow. We note that crashes may themselves be a downstream consequence of emotionally induced speeding rather than an independent confounder; we therefore present results both with and without this control to assess its influence. $\delta_{it}$ includes a set of time-varying control variables such as weather conditions (e.g., temperature extremes, precipitation, visibility), holiday indicators, and (for NBA games) day-of-week and month or (for NFL games) week-of-season fixed effects and season dummies. Standard errors are clustered at the TMC-code level to account for serial correlation within road segments over time.

The term $f(p_{it}, y_{it}; \lambda)$ represents a function of pre-game expectations and actual game outcomes. The pre-game expectation $p_{it}$ is defined using the point spread from betting markets. Based on these spreads, we classify each game into one of three mutually exclusive pre-game expectation categories (with the home team as the unit of observation/point of reference): predicted win, predicted close game, and predicted loss. These thresholds vary by sport as described in Section~\ref{sec:sportsdata}, with symmetric $\pm$4-point cutoffs applied to NFL games (following \citet{Card2011}) and asymmetric cutoffs of $-6.75$ and $+4.75$ for NBA games (following \citet{Cardazzi2024}). The realized outcome of the game, $y_{it}$, is coded as a win or loss for the home team. Following \citet{Card2011}, interactions between the pre-game expectation $p_{it}$ and the actual game outcome $y_{it}$ allow us to isolate the behavioral effects of different expectation--outcome combinations, corresponding to the suspense--valence profiles described in Section~\ref{sec:framework}. As such, we can rewrite $f(\cdot)$ as:

\begin{equation}
\begin{aligned}
f(\cdot) = & \lambda_1 \text{predicted win}_{it} + \lambda_2 (\text{predicted win}_{it} \times \text{lose}_{it}) \\
& + \lambda_3 \text{predicted close}_{it} + \lambda_4 (\text{predicted close}_{it} \times \text{lose}_{it}) \\
& + \lambda_5 \text{predicted loss}_{it} + \lambda_6 (\text{predicted loss}_{it} \times \text{win}_{it})
\end{aligned}
\label{eq:function}
\end{equation}

The coefficients $\lambda_1$, $\lambda_3$, and $\lambda_5$ capture the average speed difference on game days relative to matched non-game-day observations for each pre-game expectation category. These coefficients are potentially interesting but less directly interpretable, since variation in the point spread may be correlated with other factors that independently affect traffic conditions \citep{Card2011}. The main coefficients of interest are $\lambda_2$, $\lambda_4$, and $\lambda_6$, which measure the differential effect of the actual outcome within each expectation category. Specifically, $\lambda_2$ captures the effect of an upset loss (surprise combined with negative valence), $\lambda_4$ captures the effect of a predicted-close loss (sustained suspense combined with negative valence), and $\lambda_6$ captures the effect of an upset win (surprise combined with positive valence). Note that within each expectation category, the interaction term compares the two possible outcomes directly: $\lambda_4$, for example, captures the speed difference between predicted-close losses and predicted-close wins, with the latter serving as the implicit reference outcome. This within-category comparison isolates the effect of outcome valence while holding the level of pre-game suspense constant.

For each 10-minute event-time window, we estimate a separate regression on the subset of observations corresponding to that window on game days and matched non-game-day observations occurring at the same time of day and day of week. Non-game-day control observations are identified via pre-constructed matching indicators in the data, which flag non-game-day observations at the same time of day and day of week as each game-day window, restricted to dates falling within the regular season window defined by the first and last home game of each season. This ensures that non-game-day control observations reflect comparable seasonal traffic conditions rather than off-season patterns. Our regressions are stratified by distance from the stadium in 1 km increments (0--5 km), and we estimate separate models for each of the 18 × 10-minute intervals (6 pre-game and 12 post-game), resulting in 90 separate models across time and space, whose coefficients are plotted in Figures~\ref{fig:nfl} and~\ref{fig:nba} to trace how effects evolve over time and distance from the stadium.

In summary, by estimating separate models for each event-time window, matching game-day observations to non-game-day observations within the regular season window, and conditioning on both environmental and temporal factors, our design provides credible within-road-segment comparisons to identify the traffic patterns that follow emotionally salient sporting events with distinct suspense--valence profiles. The identifying assumption is that, conditional on the pre-game point spread, the actual game outcome is as good as random \citep{Card2011}, so that the interaction coefficients $\lambda_2$, $\lambda_4$, and $\lambda_6$ capture the causal effects of each expectation--outcome profile on post-game driving speed. A positive estimate on these terms reflects higher post-game speeds following that outcome profile relative to the base effect of the pre-game expectation category.

To complement our speed analysis, we also estimate models using crash counts as the dependent variable to assess whether emotional effects on driving speed translate into observable differences in collision frequency. Because crashes are sparse count data, we estimate Poisson fixed-effects regressions analogous to the main specification, using the same temporal windows (60 minutes pre-game, 60 and 120 minutes post-game) but aggregated across the full 0--5 km radius rather than disaggregated by distance bands. Formally, we estimate:

\begin{equation}
\log(\text{Crash}_{it}) = \alpha_i + f(p_{it}, y_{it}; \lambda) + \delta_{it} + \varepsilon_{it}
\label{eq:crash}
\end{equation}

where $\text{Crash}_{it}$ denotes the number of reported crashes on segment $i$ at time $t$. The functional form $f(\cdot)$ mirrors Equation~\ref{eq:function}, and the fixed effects and control variables are identical. The detailed estimates are presented in Appendix Tables~\ref{tab:nfl_crash} and~\ref{tab:nba_crash}.

\section{Empirical findings}
\label{sec:results}

\subsection{Baseline results}
\begin{figure}[h!t]
\centering
\refstepcounter{figure}
\label{fig:nfl}
\captionsetup{justification=centering, font=small, labelfont=bf, labelsep=colon}
\caption*{\textbf{Figure \thefigure}\\[4pt]Effects of expected and unexpected NFL game outcomes on vehicle speed near stadiums}
\vspace{4pt}
\includegraphics[width=\textwidth]{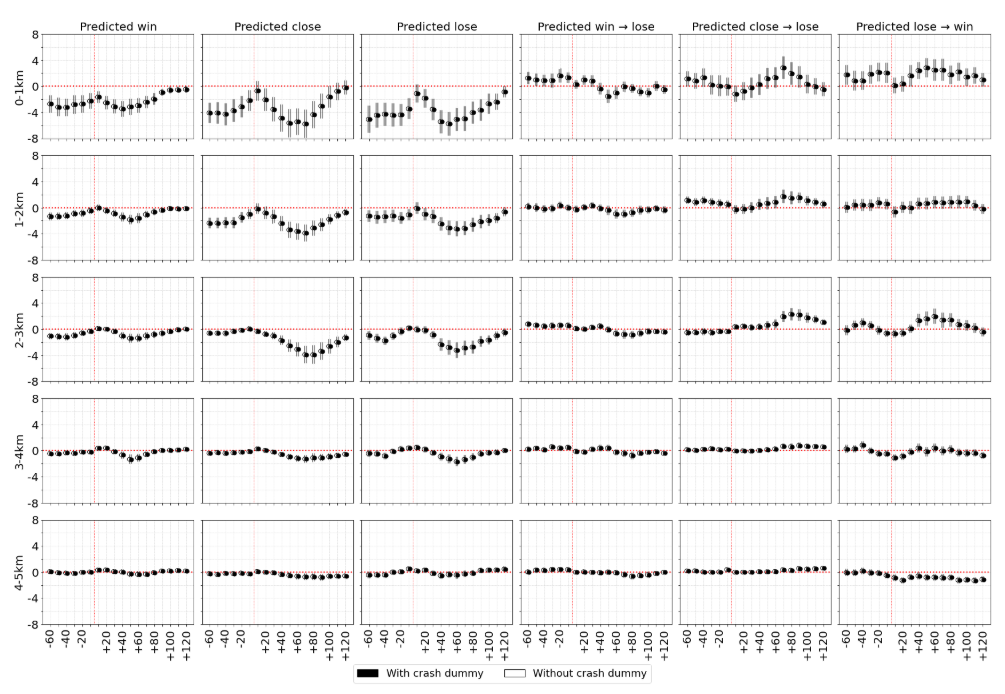}
\noindent
\begin{minipage}{\textwidth}
  \vspace{8pt}
  \begin{spacing}{0.9}
  \scriptsize
  \noindent\textit{Notes:} Predicted win indicates a point spread of --4 or less (negative spreads indicate the number of points a team is expected to win by); predicted close indicates a point spread between --4 and +4 exclusive; and predicted loss indicates a spread of +4 or more. The sample is limited to Sundays during the regular-season home games for Florida's three NFL teams---the Miami Dolphins, Tampa Bay Buccaneers, and Jacksonville Jaguars---across the 2015--2019 seasons. Due to missing traffic data, the 2018 and 2019 Miami Dolphins seasons are excluded, resulting in a total of 84 Sunday home games with start times between 1:00 PM and 4:25 PM Eastern Time. Regressions include TMC-code fixed effects, as well as week-of-season time fixed effects and season dummies to control for temporal variation. Standard errors are clustered at the TMC-code level. Average vehicle speed represents observed speed across road links within 5 km of each stadium, broken up into 1-km distance bands, measured in 10-minute intervals spanning 60 minutes before to 120 minutes after each game. Average attendance across the three Florida NFL teams during the sample period was approximately 60,847 per game (Jacksonville Jaguars: 62,702; Miami Dolphins: 65,168; Tampa Bay Buccaneers: 56,560).
  \end{spacing}
\end{minipage}
\end{figure}

Figure~\ref{fig:nfl} presents the estimated effects of NFL game outcomes on average vehicle speed across 18 $\times$ 10-minute intervals (6 pre-game and 12 post-game). Each of the 30 subplots corresponds to a unique combination of spatial proximity to the stadium (5 radii) as well as pre-game expectations and actual game outcomes (6 variables). For each distance-time cell, we estimate a fixed-effects panel regression with entity-clustered standard errors as specified in Section~\ref{sec:empirical}. We acknowledge that residuals may exhibit both temporal dependence within road segments and spatial dependence across nearby segments; the robustness of our inference to alternative clustering approaches is assessed in Section~\ref{sec:se_robust}. Each regression is estimated twice, once including a crash control dummy and once without it, yielding a total of 180 regressions (5 radii $\times$ 18 time intervals $\times$ 2 (crash/no crash), see Table~\ref{tab:nfl_table} for concrete numbers).

\vspace{0pt}

The red vertical line in each subplot denotes the scheduled start/end of the game, anchoring all observations in event time. The x-axis spans 60 minutes prior to the game start through 120 minutes after the game end. The y-axis reflects the estimated change in speed (miles/hour) either relative to matched non-game-day observations occurring at the same clock time and same day of the week (columns 1--3) or games with alternative outcomes within the same expectation category (columns 4--6). In other words: while all subplots in columns 1--3 reflect common traffic patterns on game days (compared to non-game days), the subplots in columns 4--6 reflect traffic patterns following the suspense--valence profiles described in Section~\ref{sec:framework}, i.e., a loss (compared to a win) when a win was predicted, a loss (compared to a win) when the game was predicted to be close, or a win (compared to a loss) when a loss was predicted. The minimal difference between estimates with (filled black circles) and without crash controls (hollow circles) suggests that crash-induced congestion does not drive any of the patterns we observe. We note, however, that the crash control is measured at the TMC-segment level and may not fully capture spillover effects to adjacent segments; we treat this as a limitation of the crash control approach. Moreover, average speed reflects aggregate traffic behavior and cannot directly distinguish individual aggressive driving from changes in traffic composition, such as a shift toward faster vehicle types or a thinning of slower vehicles after the game.

To start with, the subplots in columns 1--3 in Figure~\ref{fig:nfl} reveal clear and consistent traffic slowdowns before and after the NFL games. Average speed declines before the game starts and begins declining within 10 minutes after the game ends, reaches a minimum roughly 40 to 70 minutes post-game, and then gradually recovers. This pattern is most evident in the 0--1 km range but diminishes with distance and is hardly visible anymore in the 4--5 km range. Importantly, even though this pattern seems to be more pronounced for predicted losses and predicted close games, it is generally evident and consistent across all three predictions. 

Figure~\ref{fig:nfl} further reveals some interesting patterns following different expectation--outcome combinations. The clearest and most robust effect emerges for predicted-close games that ended in a loss (column 5) --- the high-suspense, negative-valence profile identified in Section~\ref{sec:framework} as theoretically most likely to generate post-game behavioral spillovers. These games exhibit significant post-game increases in vehicle speed within the 0--3 km range, with the largest changes observed during the first hour after the game ends. Average speed following predicted-close losses increases by up to 3 mph relative to predicted-close wins --- a difference several times larger than the average game day versus non-game day speed differential reported in Table~\ref{tab:descriptive_stats}, underscoring the behavioral and economic significance of this effect. This combination of sustained suspense and negative valence appears to translate into measurably higher post-game vehicle speeds, which is suggestive of more aggressive driving behavior, though we cannot test this directly. The spatial decay we see between 4--5 km suggests that these emotional spillovers are highly localized and likely tied to the immediate post-game context, rather than generalized driving behavior across the city.

While we do observe some smaller though less precise positive effects for upset wins, consistent with the surprise--positive valence profile (column 6), the coefficients for upset losses, which combine surprise with negative valence, (column 4) mainly oscillate around zero. In the robustness checks we try to explore whether this can be traced back to real differences in observed driving behavior after different suspense--valence profiles or whether this is just an artifact caused by the small number of observations.\footnote{Recall, while we observe 58 games that were predicted to be close, we only observe 16 (10) games that were predicted wins (losses) out of which just 5 (6) ended in an upset loss (win).}

\begin{figure}[ht]
\centering
\refstepcounter{figure}
\label{fig:nba}
\captionsetup{justification=centering, font=small, labelfont=bf, labelsep=colon}
\caption*{\textbf{Figure \thefigure}\\[4pt]Effects of expected and unexpected NBA game outcomes on vehicle speed near arenas}
\vspace{4pt}
\includegraphics[width=\textwidth]{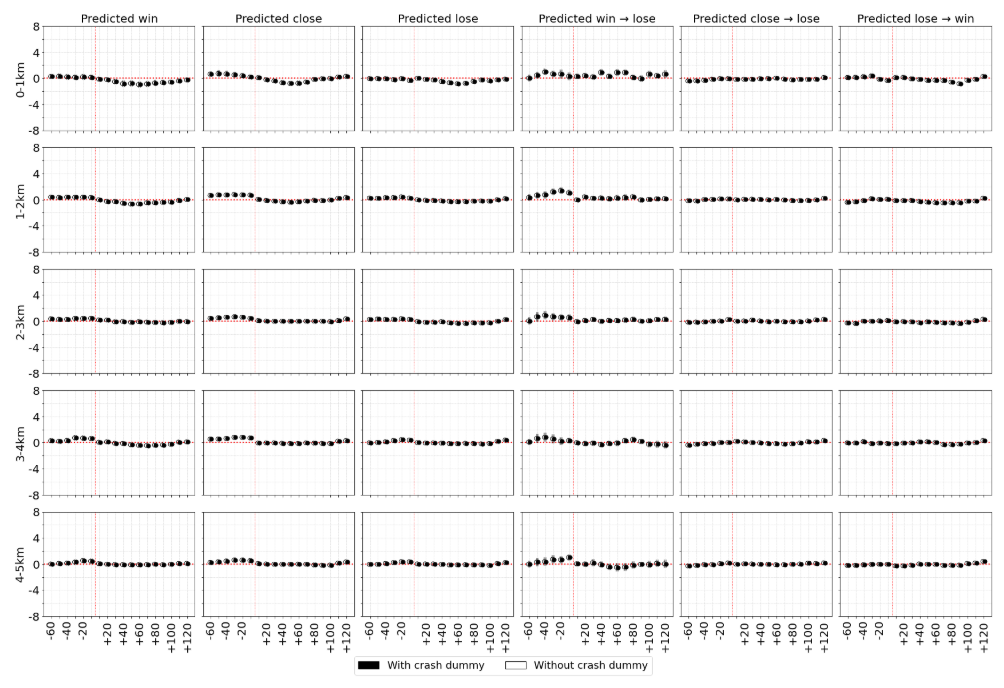}
\noindent
\begin{minipage}{\textwidth}
  \vspace{8pt}
  \begin{spacing}{0.9}
  \scriptsize
  \noindent\textit{Notes:} Predicted win indicates a point spread of --6.75 or less (negative spreads indicate the number of points a team is expected to win by); predicted close indicates a point spread between --6.75 and +4.75 exclusive; and predicted loss indicates a spread of +4.75 or more. The sample is limited to regular-season NBA games between the 2015/16 and 2018/19 seasons, encompassing four total seasons. The analysis includes home games played by Florida's two NBA teams---the Miami Heat and the Orlando Magic---with scheduled tip-off times of 6:00 PM or later. Regressions include TMC-code fixed effects, as well as month time fixed effects, season dummies, and day-of-week dummies to control for temporal variation. Standard errors are clustered at the TMC-code level. Average vehicle speed represents observed speed across road links within 5 km of each arena, broken up into 1-km distance bands, measured in 10-minute intervals spanning 60 minutes before to 120 minutes after each game. Average attendance across the two Florida NBA teams during the sample period was approximately 18,651 per game (Miami Heat: 19,665; Orlando Magic: 17,651).
  \end{spacing}
\end{minipage}
\end{figure}

Figure~\ref{fig:nba} shows the results for NBA games (see Table~\ref{tab:nba_table} for full results), estimating changes in average vehicle speed around home games played by Florida teams. Starting with columns 1--3, we observe a consistent---though compared to NFL games much smaller---decline in traffic speed beginning approximately 10 minutes after the game ends, bottoming out between 40 and 70 minutes post-game, and then gradually returning toward baseline. This pattern is consistent with standard post-game traffic congestion and is most pronounced in the 0--1 km range, attenuating with distance from the arena. In contrast to our findings for the NFL, however, all estimates for the suspense--valence profiles from NBA games (columns 4--6) mainly oscillate around zero and lack any systematic post-game movement. Again, the inclusion or exclusion of crash controls does not materially alter this finding. As such, and in contrast to our findings for the NFL games, we find no evidence that any of the expectation--outcome profiles in NBA games affect driving behavior as measured by average vehicle speed. While post-game traffic slowdowns are visible and local in nature, neither the high-suspense negative-valence profile nor the surprise--positive or surprise--negative valence profiles appear to generate measurable behavioral spillovers into the driving environment.

\subsection{Robustness checks}
\label{sec:robust}

To assess the stability of our findings for the NFL, we conduct a series of tests confirming that the observed post-game speed increases are not driven by specific game start times, incomplete team coverage, or different modeling choices. We first examine the sensitivity of our inference to alternative standard error specifications, including two-way clustering and a wild cluster bootstrap procedure. We then assess robustness to kickoff time restrictions, team coverage, alternative spread thresholds, and controls for belief updating. We close this section by providing some counterfactual (during game) evidence, followed by two placebo tests: a randomization placebo in which win/loss outcome labels are randomly reassigned within each spread category, and a pseudo game-day placebo where real game labels are assigned to non-game Sundays. Together, these exercises demonstrate that the documented emotional-driving effects are robust across specifications and samples.

\subsubsection{Alternative standard error specifications}
\label{sec:se_robust}

Our baseline specification uses entity-clustered standard errors at the TMC-code level, which account for serial correlation within road segments over time. To assess whether inference is sensitive to this choice, we consider two additional clustering approaches, as well as a wild cluster bootstrap procedure.

First, we adopt two-way clustered standard errors that cluster simultaneously at the TMC-code (entity) level and the season-week (time) level, following \citet{Cameron2011}. We implement this under two fixed-effects structures: (i) the baseline week-of-season and season fixed effects, and (ii) a more demanding specification using season-week fixed effects (84 unique time periods). Appendix Figures~\ref{fig:nfl_twoway_regfe} and~\ref{fig:nfl_twoway_weekseasonfe} present the results. Under both specifications, the core finding of elevated post-game speed following predicted-close losses remains robust, with significant effects concentrated in the 2--3 km distance band during the 70--110 minutes post-game window. The upset win effect remains significant under two-way clustering with the baseline fixed effects, though it becomes imprecise under the more demanding season-week fixed-effects specification, suggesting that this effect is more sensitive to the time fixed-effects structure than the predicted-close loss effect.

Second, we implement a wild cluster bootstrap \citep{Cameron2015} to address potential finite-sample concerns with cluster-robust inference. Following the recommendation of \citet{Roodman2019}, we use Rademacher weights with the null imposed (WCR bootstrap, type ``11''). The bootstrap p-values, presented in Appendix Table~\ref{tab:wildboot}, confirm the significance of the predicted-close loss effect at the 1\% level across the key post-game windows (+70 to +100 minutes) and the 2--3 km distance band. Bootstrap p-values for the upset win coefficient are at or below 0.001 across most post-game windows, providing strong confirmation that the surprise--positive valence effect on driving speed is statistically robust under finite-sample inference.

Together, these exercises confirm that the main findings are not an artifact of the standard error specification. The predicted close $\times$ loss effect, combining sustained suspense with negative valence, is consistently significant under all three alternative inference procedures. The upset win effect holds under entity clustering and wild bootstrap but is less stable under the more conservative two-way clustering with season-week fixed effects.

\subsubsection{Kickoff time}
\label{sec:sampletiming}

We also test whether game start time influences the results we find. Restricting the sample to NFL games with 1:00 PM kickoffs eliminates variation arising from later start times (4:05 PM or 4:25 PM), which could be affected by differing travel conditions, lighting, or crowd dispersal patterns. The resulting estimates, presented in Appendix Figure~\ref{fig:nfl_1pm}, are nearly identical to those in the baseline model (Figure~\ref{fig:nfl}). Overall, our main findings remain: average post-game speed increases significantly following predicted-close losses, the high-suspense, negative-valence profile, with similar magnitudes and timing across all distance bands as in our main specification. Moreover, the effects for upset wins, combining surprise with positive valence, even become more pronounced.

\subsubsection{Team coverage}
\label{sec:nodolphins}

Next, we evaluate whether incomplete traffic data for the Miami Dolphins during the 2018 and 2019 seasons influence the results. We re-estimate the regressions excluding all Dolphins home games, leaving only Tampa Bay Buccaneers and Jacksonville Jaguars games in the sample. Appendix Figure~\ref{fig:nfl_nodolphins} shows that the estimated patterns remain virtually unchanged relative to the full-sample results. Post-game speeds still rise significantly following predicted-close losses and upset wins, with no notable changes in magnitude or timing across distance bands. Overall, these results confirm that incomplete data for the Miami Dolphins do not drive the baseline findings for the NFL.

\subsubsection{Spread thresholds}
\label{sec:spreadthresholds}

To assess whether our findings depend on the definition of pre-game expectations, we replicate our analysis using alternative point-spread cutoffs of $\pm3$ and $\pm5$ points (instead of $\pm4$). Table~\ref{tab:nfl_spread_bins} provides an overview of how the distribution of games changes under different thresholds highlighting how narrower (wider) cutoffs identify fewer (more) close games and more (fewer) wins and losses respectively. Importantly, there is some asymmetry here, i.e., while decreasing the threshold from $\pm4$ to $\pm3$ causes 26 games to be classified differently, increasing the threshold from $\pm4$ to $\pm5$ causes only 5 games to be classified differently.

\begin{table}[ht]
\centering
\refstepcounter{table}
\label{tab:nfl_spread_bins}
\captionsetup{justification=centering, font=small, labelfont=bf, labelsep=colon}
\caption*{\textbf{Table \thetable}\\[4pt]Game counts by spread threshold and outcome category}
\vspace{6pt}
\scriptsize
\setlength{\tabcolsep}{0pt}
\renewcommand{\arraystretch}{1}
\begin{tabular*}{\textwidth}{@{\extracolsep{\fill}}lccccccc@{}}
\toprule
Version & \makecell{Predicted\\win} & \makecell{Predicted\\close} & \makecell{Predicted\\loss} & \makecell{Predicted win\\$\times$ lose} & \makecell{Predicted close\\$\times$ lose} & \makecell{Predicted loss\\$\times$ win} \\
\midrule
Spread $\pm2$ & 40 & 17 & 27 & 18 & 10 & 12 \\
Spread $\pm3$ & 31 & 32 & 21 & 12 & 18 & 8 \\
Baseline $\pm4$ & 16 & 58 & 10 & 5 & 34 & 6 \\
Spread $\pm5$ & 13 & 63 & 8 & 3 & 37 & 5 \\
Spread $\pm6$ & 10 & 69 & 5 & 2 & 39 & 3 \\
\bottomrule
\end{tabular*}
\noindent
\begin{minipage}{\textwidth}
  \vspace{10pt}
  \begin{spacing}{0.9}
  \scriptsize
  \noindent\textit{Notes:} Counts reflect the number of NFL home games in each predicted outcome and interaction category under alternative spread thresholds. Narrower cutoffs (e.g., $\pm2$, $\pm3$) reduce the number of close games and increase predicted wins and losses; wider cutoffs (e.g., $\pm5$, $\pm6$) do the opposite. For example, under the $\pm6$ threshold, any game with a point spread at or below --6 is classified as a predicted win, any at or above +6 as a predicted loss, and those between --6 and +6 (exclusive) as predicted close contests. The sample is limited to Sundays during the regular-season home games for Florida's three NFL teams---the Miami Dolphins, Tampa Bay Buccaneers, and Jacksonville Jaguars---across the 2015--2019 seasons. Due to missing traffic data, the 2018 and 2019 Miami Dolphins seasons are excluded, resulting in a total of 84 Sunday home games with start times between 1:00 PM and 4:25 PM Eastern Time.
  \end{spacing}
\end{minipage}
\end{table}

Overall, we still observe positive effects on average speed following predicted-close losses using a cut-off at $\pm3$, though these are smaller than in the baseline. The upset win effect---combining surprise with positive valence---however vanishes under this narrower threshold (if anything, there is even a small but negative effect within the 1--3 km range, see Figure~\ref{fig:nfl_spread_3}). This is consistent with the narrower threshold reclassifying many games previously identified as predicted wins or losses into the predicted-close category, diluting the suspense--valence contrast. In contrast, using a cut-off at $\pm5$ we find very similar results to our main specification with both the predicted-close loss effect and the upset win effect being even more pronounced (see Figure~\ref{fig:nfl_spread_5}).

\subsubsection{Updating beliefs}
\label{sec:halftimeupdate}

In order to test whether updating beliefs is relevant in our setting, we consider score differentials at halftime as proposed by \citet{Card2011}. More precisely, we include halftime status as a control variable to isolate its influence relative to pregame expectations. A second specification focuses directly on halftime (instead of before the game) predictions, examining how post-game traffic varies considering the home team's position at halftime and the final result. As discussed in Section~\ref{sec:framework}, we proxy for sustained suspense using the pre-game point spread under the assumption that games predicted to be close are more likely to remain competitive throughout. The halftime specifications below partially test this assumption by examining whether within-game score dynamics alter the main findings.

In line with \citet{Card2011}, we find little evidence of systematic belief updating based on halftime information affecting our results. In the control specification (Figure~\ref{fig:nfl_halftimecontrol}), the patterns for predicted-close losses---the high-suspense, negative-valence profile---and upset wins largely mirror the results in Figure~\ref{fig:nfl}, though the effects are slightly less precise. In the halftime prediction specification (Figure~\ref{fig:nfl_halftimeupdate}), the coefficients for games predicted to be close at halftime but that ultimately ended in a loss show no consistent pattern, and any negative deviations in post-game speeds are small and confined to the immediate stadium area (0--1 km). Similarly, games in which the home team was predicted to lose at halftime but ultimately won show at most a minor and imprecise positive response within the 0--1 km range. Together, these results suggest that even though fans are likely to adjust their expectations as games unfold, such belief updating during games plays only a minor role in shaping post-game driving speeds. This is consistent with our conceptual framework, which emphasizes pre-game suspense and final outcome valence as the primary drivers of post-game behavioral responses, rather than within-game dynamics. We also examined whether games featuring a fourth-quarter lead change --- for instance, teams winning going into the fourth quarter but ultimately losing, or vice versa --- generate distinct post-game driving patterns. However, with only 9 such heartbreak losses and 6 late comeback wins among our 84 games, the sample is too small to support reliable inference for these categories. We therefore flag more granular within-game belief updating, including quarter-by-quarter score dynamics and late lead changes, as a promising avenue for future research, particularly in settings with larger game samples.

\subsubsection{Counterfactual analysis}
\label{sec:duringgame}

To relax the possible concern that any traffic patterns we observe are simply game-day related rather than driven by the suspense--valence profiles described in Section~\ref{sec:framework}, we test during-game patterns in a counterfactual setup. More precisely, we construct a panel of 10-minute bins beginning at game start and extending to 170 minutes post-kickoff, corresponding to the duration of the shortest game in the sample. Figure~\ref{fig:nfl_duringgame} displays the estimated changes in speed by expectation--outcome combination and distance band. The interpretation mirrors our pre/post-game specification since each time bin is compared to a matched non-gameday bin of the same time-of-day and day of week (columns 1--3) or to games with alternative outcomes within the same expectation category (columns 4--6).

As can be seen in Figure~\ref{fig:nfl_duringgame}, traffic speed remains rather stable during games. Whether the home team was predicted to win, lose, or play in a close contest, and regardless of the final outcome, we observe no meaningful change in driving speed relative to non-gamedays (or alternative outcomes). The only exception concerns upset wins, the surprise--positive valence profile, for which we observe mildly reduced average speed. Overall, however, these results reinforce our central claim that the emotional context generated by the final outcome andthe pre-game expectation of a close contest, rather than the mere occurrence of a game, may drive the changes in traffic behavior we observe.

\subsubsection{Randomization placebo}
\label{sec:placebo_random}

To verify that our main findings are not an artifact of the data structure or estimation procedure, we implement a randomization placebo test. We run our full regression model 150 times, each time randomly shuffling the win/loss outcome labels within each spread category before estimation --- keeping spread category assignments fixed but randomizing the actual outcome within each category. Everything else remains identical: the same games, same dates, same stadiums, same traffic data, same weather controls, and same fixed effects.

Appendix Figure~\ref{fig:nfl_placebo_random} presents the average coefficient across the 150 draws for each time window and distance band, along with the 2.5 and 97.5 percentile range of the placebo coefficients --- reflecting what we would expect to find by chance under random outcome assignment. For all three interaction terms the placebo mean is essentially zero across all time windows and distance bands, with no systematic post-game pattern. The percentile bands for the predicted loss $\times$ win interaction are somewhat wider, reflecting the small number of predicted loss games in the sample (only 10), though the mean remains close to zero throughout. The real coefficient for predicted-close losses at the key post-game windows sits well outside the 97.5 percentile of the placebo coefficients, confirming that the main finding cannot be replicated by random outcome assignment.

\subsubsection{Pseudo game-day placebo}
\label{sec:placebo_pseudogame}

As a further placebo test, we examine whether the post-game speed patterns we observe could be driven by normal Sunday afternoon traffic patterns, rather than by the actual occurrence of an NFL game and accompanying emotions. We restrict the sample to non-game Sundays only and construct a set of pseudo game days by randomly assigning each of the 84 real games to a non-game Sunday from the same team's stadium area, without replacement. Each assigned date inherits the real game's spread category and outcome label, and the existing matched time-window indicators, which flag observations occurring at the same time of day as a real game's pre- and post-game windows, are used as event-time dummies for the pseudo game days. The remaining non-game Sundays serve as the control group. The same regression specification as the main model is then estimated on this pseudo game-day dataset. As with the randomization placebo in Section~\ref{sec:placebo_random}, we repeat this procedure 150 times, each time with a new random assignment of real games to non-game Sundays.

Across the three teams, a total of 135 non-game Sundays are available for assignment: 52 for the Jacksonville Jaguars, 32 for the Miami Dolphins, and 51 for the Tampa Bay Buccaneers. After assigning one non-game Sunday to each of the 84 real games, 51 non-game Sundays remain as control observations (Jacksonville: 20, Miami: 13, Tampa Bay: 18).

As shown in Appendix Figure~\ref{fig:nfl_placebo_pseudogame}, the average coefficient across the 150 draws is, across all six outcome profiles, all five distance bands, and all 18 time windows, essentially zero, with no systematic pre- or post-game pattern, and the 2.5th to 97.5th percentile range of the placebo coefficients remains narrow and centered around zero throughout. This confirms that the post-game speed increases documented in the main analysis are not driven by normal Sunday traffic dynamics, time-of-day effects, or the event-time window construction, but are instead tied to the actual occurrence of an NFL game and its emotional outcome.

\subsection{Crash analysis}
\label{sec:crash}

Due to crash events being quite rare, we estimate Poisson fixed-effects regressions using crash counts as the dependent variable on the aggregated 0--5 km sample as well as split only into three intervals (60 minutes pre-game, 60 and 120 minutes post-game). Table~\ref{tab:nfl_crash_main} presents the results for the NFL, for gamedays as well as splitting gamedays into wins and losses. Before discussing the results, we note that crash counts in our sample are low --- 17 crashes in the pre-game window and 40 in the post-game 60-minute window across all gamedays in the 2015--2019 sample --- so the Poisson coefficients, which are expressed as log changes, should be interpreted with appropriate caution. We report percentage changes derived from the exponentiated coefficients as indicative of the direction and approximate magnitude of effects, rather than as precise estimates.

Turning to the results, Table~\ref{tab:nfl_crash_main} reveals two notable patterns. First, pre-game crash counts are already elevated on gamedays relative to matched non-gamedays. The gameday coefficient in the Pregame\_60 window is positive and statistically significant (estimate = 0.933, p = 0.008), corresponding to approximately 154\% more crashes on gamedays relative to non-gamedays in the hour before kickoff. This likely reflects elevated pre-game travel and fan activity around stadiums rather than emotional driving, and represents a new finding relative to earlier versions of this analysis.

Second, our estimates reveal a clear post-game increase in crashes for both time windows. In the Postgame\_60 window, gameday crashes are approximately 404\% more frequent relative to matched non-gamedays (estimate = 1.618, p = 0.000). This increase is somewhat larger following losses (approximately 505\% more crashes, estimate = 1.798, p = 0.000) than following wins (approximately 408\% more crashes, estimate = 1.624, p = 0.000). In the Postgame\_120 window, the pattern persists: gameday crashes remain significantly elevated (approximately 243\% more frequent, estimate = 1.232, p = 0.000), again with a larger effect following losses (approximately 353\% more crashes, estimate = 1.510, p = 0.000) than wins (approximately 181\% more crashes, estimate = 1.032, p = 0.015). While these percentage increases are large in relative terms, they should be interpreted in the context of the low absolute crash counts involved.

\begin{table}[htbp]
\centering
\refstepcounter{table}
\label{tab:nfl_crash_main}
\captionsetup{justification=centering, font=small, labelfont=bf, labelsep=colon}
\caption*{\textbf{Table \thetable}\\[4pt]Poisson fixed-effects regressions of NFL crash counts by win/loss outcome}

\vspace{6pt}

\scriptsize
\setlength{\tabcolsep}{8pt}
\renewcommand{\arraystretch}{1}
\begin{tabular*}{\textwidth}{@{\extracolsep{\fill}}lccccc@{}}
\toprule
\textbf{Period} & \textbf{Number of crashes} & \textbf{Estimate} & \textbf{\textit{p}-value} & \textbf{95\% CI Low} & \textbf{95\% CI High} \\
\midrule
\textbf{Pregame\_60} & & & & & \\[2pt]
Gamedays (non-gamedays) & 17 (33) & 0.933 & 0.008 &  0.249 & 1.616 \\
[4pt]
\textbf{Postgame\_60} & & & & & \\[2pt]
Gamedays (non-gamedays) & 40 (52) & 1.618 & 0.000 &  1.159 & 2.077 \\
Wins & 18 &  1.624 & 0.000 &  0.972 & 2.275 \\
Losses & 22 &  1.798 & 0.000 &  1.284 & 2.313 \\
[4pt]
\textbf{Postgame\_120} & & & & & \\[2pt]
Gamedays (non-gamedays) & 21 (41) & 1.232 & 0.000 &  0.662 & 1.803 \\
Wins & 9  &  1.032 & 0.015 &  0.196 & 1.867 \\
Losses & 12 &  1.510 & 0.000 &  0.776 & 2.243 \\
[4pt]
\bottomrule
\end{tabular*}

\noindent
\begin{minipage}{\textwidth}
  \vspace{10pt}
  \begin{spacing}{0.9}
  \scriptsize
  \noindent\textit{Notes:} The dependent variable is the crash count per TMC segment and 60-minute time interval. Crash counts reported in this table reflect the total number of unique crash events aggregated across all observation windows in the 2015--2019 sample (i.e., all gamedays and all corresponding matched non-gameday control periods), rather than averages per gameday or per non-gameday. Estimates are from Poisson fixed-effects regressions on the aggregated 0--5 km sample. The sample is limited to Sundays during the regular-season home games for Florida's three NFL teams---the Miami Dolphins, Tampa Bay Buccaneers, and Jacksonville Jaguars---across the 2015--2019 seasons. Due to missing traffic data, the 2018 and 2019 Miami Dolphins seasons are excluded, resulting in a total of 84 Sunday home games with start times between 1:00 PM and 4:25 PM Eastern Time. Poisson regressions include TMC-code fixed effects, as well as week-of-season time fixed effects and season dummies to control for temporal variation. Standard errors are clustered at the TMC-code level. Coefficients are reported as log changes; positive values indicate higher crash frequencies relative to matched non-gameday periods.
  \end{spacing}
\end{minipage}
\end{table}

Due to the very limited number of crash observations, further splitting the sample does not yield any consistent patterns (see Table~\ref{tab:nfl_crash} in the Appendix). Still, it is worth noting that elevated post-game speeds around NFL stadiums coincide with higher crash counts on gamedays relative to non-gamedays, particularly after losses. This pattern is consistent with the speed analysis: the high-suspense, negative-valence profile of predicted-close losses generates the largest post-game speed increases, and the crash analysis suggests that these speed increases may translate into elevated collision risk. Together, the speed and crash findings paint a coherent picture of emotionally induced behavioral spillovers from NFL games into post-game traffic.\footnote{Crash count regressions for the NBA do not indicate any significant effects (see Table~\ref{tab:nba_crash} in the Appendix).}

\section{Discussion}
\label{sec:discussion}

Our analysis provides robust evidence of emotional driving after NFL games, concentrated in the high-suspense, negative-valence profile: games predicted to be close that ended in disappointing home-team losses. Average vehicle speed rises by up to 3 mph relative to predicted-close games that ended in a win in the first post-game hour. As shown in Table~\ref{tab:descriptive_stats}, the average difference between game day and non-game day speeds is less than 1 mph across all distance bands, meaning the 3 mph effect following predicted-close losses is several times larger than the typical game day deviation --- underscoring the behavioral and economic significance of this finding. The spatial decay we see between 4--5 km suggests that these emotional spillovers are highly localized and likely tied to the immediate post-game context, rather than generalized driving behavior across the city. We also find some smaller---though less precise---positive effects on average speed following upset wins --- the surprise--positive valence profile --- while we do not observe any consistent patterns following upset losses --- the surprise--negative valence profile. Importantly, however, sample size restrictions limit the extent of our investigation for both upset wins and upset losses. As such, despite our non-findings here, we refrain from concluding that upset losses in the NFL generally do not unfold any such effects on average vehicle speed.

In contrast to these findings for the NFL, NBA games---for which we also observe a considerable number of upset wins and losses in our data---show no systematic relationship between expectation--outcome profiles and post-game traffic behavior. While modest traffic slowdowns occur near arenas, these reflect standard congestion patterns rather than emotional responses. One may speculate that the NBA's higher game frequency (41 home games in the NBA compared to 8 in the NFL) and scoring volatility may attenuate the emotional salience of any single game outcome. By contrast, each NFL game carries greater weight in the season, and the sport's high-impact, stop-start rhythm fosters more intense emotional investment and sharper post-game emotions. For our setting, however, the most notable difference between NFL and NBA games certainly relates to the number of spectators. Average attendance is much smaller in the NBA (approximately 18,651 per game across the two Florida teams) compared to the NFL (approximately 60,847 per game across the three Florida teams), reducing overall crowd-related traffic pressure and the probability to observe a sufficiently large portion of spectators among the drivers in a 5-kilometer radius after the games. Moreover, NBA games in Florida take place in denser urban environments with limited road capacity and lower average vehicle speeds, which may dampen opportunities for speeding. In summary, we caution against interpreting this contrast as evidence that NBA games generate weaker emotional responses among fans per se. The absence of detectable effects in the NBA likely reflects structural differences in attendance, venue location, and traffic environment rather than differences in the emotional intensity experienced by individual fans.

Overall, the spatial and temporal concentration of the observed effects for NFL games supports an interpretation of transient emotional responses. Speed increases are largest within 1 km of the stadium, concentrated in the first hour after games, and dissipate beyond 3 km---consistent with the short-lived nature of emotional arousal and its immediate influence on cognition and risk perception. Within the expectations-based reference framework \citep{Koszegi2006}, predicted-close contests generate strong pre-game expectations on both sides of the reference point, and a narrow home-team loss represents the negative resolution of a high-suspense event --- combining sustained arousal with disappointing valence in a way that lopsided losses do not. The \citet{Yerkes1908} law, which suggests that performance deteriorates at high arousal levels, might help explain why we observe the largest behavioral response in these close, disappointing losses rather than in more lopsided contests. Moreover, this pattern aligns with experimental evidence that heightened emotional states---especially anger, frustration, or stress---reduce reaction times and increase risk-taking tendencies \citep[e.g.,][]{Deffenbacher2003, Jeon2014, Zhang2020, Zhang2022}. Close losses, in particular, can provoke these responses by combining high physiological activation with negative valence, thereby increasing the likelihood of aggressive or inattentive driving after the event. The timing, spatial decay, robustness checks, and absence of during-game effects are all consistent with an affective interpretation of the post-game speed increases. However, we acknowledge that the data cannot directly observe individual emotional states, and we cannot rule out all event-management and crowd-dispersal mechanisms as contributing factors; this will be discussed more in depth in Section 5.1.

Notably, \citet{Wood2011} find that close, competitive games increase traffic fatalities near the winning team's location, likely reflecting celebratory arousal and risk-taking among fans of the winning team. Our results highlight a contrasting yet complementary pattern: while \citet{Wood2011} document elevated driving risk among winning fans, we find that losing fans of close, high-suspense games also exhibit measurably riskier driving behavior through frustration and negative arousal. Together, these findings suggest that high emotional intensity in close sporting events can affect post-game driving behavior through both positive and negative affective channels, regardless of which side of the outcome fans are on. 

The effect we find is a behaviorally meaningful increase given the density of post-game traffic and the safety risks of even small deviations from speed limits. In many jurisdictions, a 3 mph deviation is enough to exceed posted speed limits and trigger enforcement, with potential implications for safety, especially near stadium exits and pedestrian crossings.\footnote{A 3 mph deviation aligns closely with the 5 km/h increase referenced in empirical studies, which show that the risk of involvement in a casualty crash can double for every 5 km/h increase in traveling speed above the limit in urban zones \citep{Kloeden1997, Kloeden2002}. Such low-level speeding (1--10 km/h over the limit) is highly prevalent and is collectively estimated to be attributable to up to 19.2\% of all casualty crashes in urban 40 km/h zones due to its high frequency \citep{Soole2023}. \citet{Wang2018} further corroborate this relationship for urban arterials specifically, finding that a 1\% increase in mean speed is associated with a 0.7\% increase in total crashes.} Taken together with our crash analysis, which shows elevated post-game crash counts on gamedays relative to matched non-gamedays, especially after losses, these findings suggest that emotional speeding is not merely cosmetic but may translate into higher collision risk.

\subsection{Limitations}
\label{sec:limitations}

Several limitations of our analysis deserve acknowledgment. First, the data do not observe individual drivers, whether a driver attended the game, whether the driver supported the home team, or the driver's emotional state. The estimated effects therefore capture changes in local traffic behavior around emotionally salient events, not direct individual-level emotional responses. In addition, average speed can change for reasons other than intentional speeding. Crowd-dispersal patterns, pedestrian flows, police traffic management, ride-share activity, parking-lot exits, and post-game celebrations may all affect observed speeds. A close win may encourage fans to linger, celebrate, or move more slowly through congested areas, while a close loss may produce faster and more synchronized departures. These mechanisms are not mutually exclusive with emotional driving, but they imply that the estimated effects should be understood as reduced-form behavioral spillovers around stadiums rather than a clean measure of individual emotional driving behavior.

Second, there may be a selection or composition issue in the surrounding traffic. On game days, some ordinary drivers may avoid the area or shift travel times, while attendees disproportionately enter the local network. As a result, non-game-day controls may not represent exactly the same underlying traffic composition. However, our pseudo game-day placebo test (Section~\ref{sec:placebo_pseudogame}) is suggestive of this not being a major concern: if non-game-day controls differed systematically in composition from game-day periods, we might expect to observe some spurious pattern even when real game labels are assigned to non-game Sundays, yet the placebo coefficients remain centered around zero with no systematic pattern. In any case, this would not invalidate the findings, but it does reinforce the idea that the estimates capture reduced-form local traffic responses to emotionally salient events, rather than a pure treatment effect on an otherwise unchanged traffic stream.

Third, some interaction cells in our NFL sample are small. While we observe 58 predicted-close games, we only observe 5 upset losses and 6 upset wins, making inference for these categories more exploratory than confirmatory. The predicted-close loss finding rests on 34 games --- a more reliable base --- but the upset win and upset loss findings should be interpreted with additional caution given the limited number of observations.

Fourth, our findings are specific to Florida NFL and NBA teams during the 2015--2019 regular seasons. Florida's climate, road network, public transportation availability, and fan culture may differ from other NFL markets. The absence of playoff games and the restriction to regular-season contests means that our findings may not generalize to higher-stakes postseason environments, where emotional intensity and attendance patterns may differ substantially.

\subsection{Policy implications}
\label{sec:policy}

Importantly, increased average vehicular speeds are predictable in time and space: they are largest within 0--3 km of stadiums and during the first post-game hour, especially following games that were expected to be close. This predictability suggests that targeted, time-bound interventions could substantially improve safety outcomes. Visible, short-term speed enforcement in the hour following close, disappointing games---through mobile radar units, automated enforcement cameras, or clearly marked police presence---could help deter impulsive speeding in these emotionally heightened periods. Municipalities might also increase pedestrian protection by deploying temporary barriers or expanding sidewalk space around stadium exits to separate walking fans from vehicle flow. Electronic roadside messaging systems --- similar to those used to alert fatigued drivers --- could be deployed near stadium exits in the post-game window, reminding drivers to slow down and exercise caution. Dynamic traffic signal systems that adjust to post-game congestion levels could further reduce stop-and-go frustration, which compounds emotional agitation behind the wheel.

Eventually, public communication and behavioral nudges can also play a preventive role. Messaging campaigns that encourage caution and remind drivers to anticipate congestion, particularly after close or high-stakes games, may reduce risky driving behavior among emotionally aroused fans. Providing post-game activities or designated waiting areas near stadiums --- such as fan zones, entertainment, or food and beverage areas --- could give emotionally aroused fans time to dissipate their emotional state before getting behind the wheel, reducing the number of emotionally affected drivers on the road immediately after games. Event organizers and local governments could coordinate these campaigns with post-game public announcements or digital signage near stadium exits. More broadly, aligning traffic management policies with the emotional salience of events may yield public safety benefits. Deploying additional traffic personnel or designating ride-share pickup zones for high-attendance or high-stakes games could reduce emotional exposure at the wheel. In the longer term, expanding or subsidizing public and shared transportation options for game days---such as discounted transit fares, dedicated shuttle routes, or ride-share vouchers---could reduce the number of emotionally affected drivers on the road. Overall, even modest shifts in driver behavior or mode choice during these predictable high-risk windows could generate meaningful safety gains for both fans and surrounding communities.


\section{Conclusions}
\label{sec:conclusion}

Using detailed traffic speed and crash data around professional sport venues in Florida, we show that NFL games predicted to be close before kickoff, generating sustained pre-game suspense, but that ultimately ended in a disappointing home-team loss produce localized increases in post-game vehicle speed. The resulting combination of sustained suspense and negative outcome valence appears to manifest as short-lived increases in risky driving behavior within 3 km of stadiums during the first post-game hour. Against the baseline of less than 1 mph average difference between game days and non-game days, the estimated 3 mph post-game speed increase following predicted-close losses is several times larger than the typical game day deviation, underscoring the behavioral significance of this effect. Further analysis suggests that the emotional context generated by the final outcome and the suspense experienced during the game, rather than congestion, environmental factors, or normal Sunday traffic patterns, drives the observed speeding.

These findings contribute to three strands of literature. They provide field evidence that expectations-based emotional cues generate behavioral spillovers into immediate real-world decisions, extending the reference-dependent preferences literature beyond laboratory settings. They identify a high-frequency behavioral mechanism, minute-level vehicle speed, linking emotionally intense sporting events to elevated public safety risks, complementing event-level crash studies. And they show that risky driving behavior can be shaped by predictable affective states generated by collective experiences, with implications for transportation policy and road safety management.

Beyond documenting this phenomenon, our findings carry important implications for public safety and event management. Because the timing and location of these high-risk emotional episodes are predictable, targeted interventions, such as visible post-game speed enforcement, dynamic traffic control, and improved pedestrian management, could effectively mitigate risk. More broadly, recognizing that emotions from collective experiences like sports can shape immediate real-world behavior underscores the value of integrating psychological and behavioral insights into traffic policy. Future research could examine whether these patterns hold in other jurisdictions, sports leagues, or playoff settings, or how specific interventions affect post-game collision rates.

Games predicted to be close before kickoff appear to create the emotional conditions most conducive to post-game speeding when the final outcome is unfavorable, combining the sustained arousal of a high-suspense contest with the frustration of a negative resolution. Anticipating and managing these predictable, emotionally charged moments offers a promising, behaviorally informed avenue for improving public safety in the aftermath of major sporting events.

\singlespacing
\bibliographystyle{apacite}
\bibliography{emotional_driving_references}

\clearpage
\renewcommand{\thefigure}{A.\arabic{figure}}
\renewcommand{\thetable}{A.\arabic{table}} 
\setcounter{figure}{0}  
\setcounter{table}{0}

\section*{Appendix}


\begin{figure}[H]
\centering
\refstepcounter{figure}
\phantomsection
\label{fig:map_bucs}
\captionsetup{justification=centering, font=small, labelfont=bf, labelsep=colon}
\caption*{\textbf{Figure  \thefigure}\\[4pt]0--5 km Radius Map for the Raymond James Stadium (Tampa Bay Buccaneers)}
\vspace{8pt}
\includegraphics[width=\textwidth]{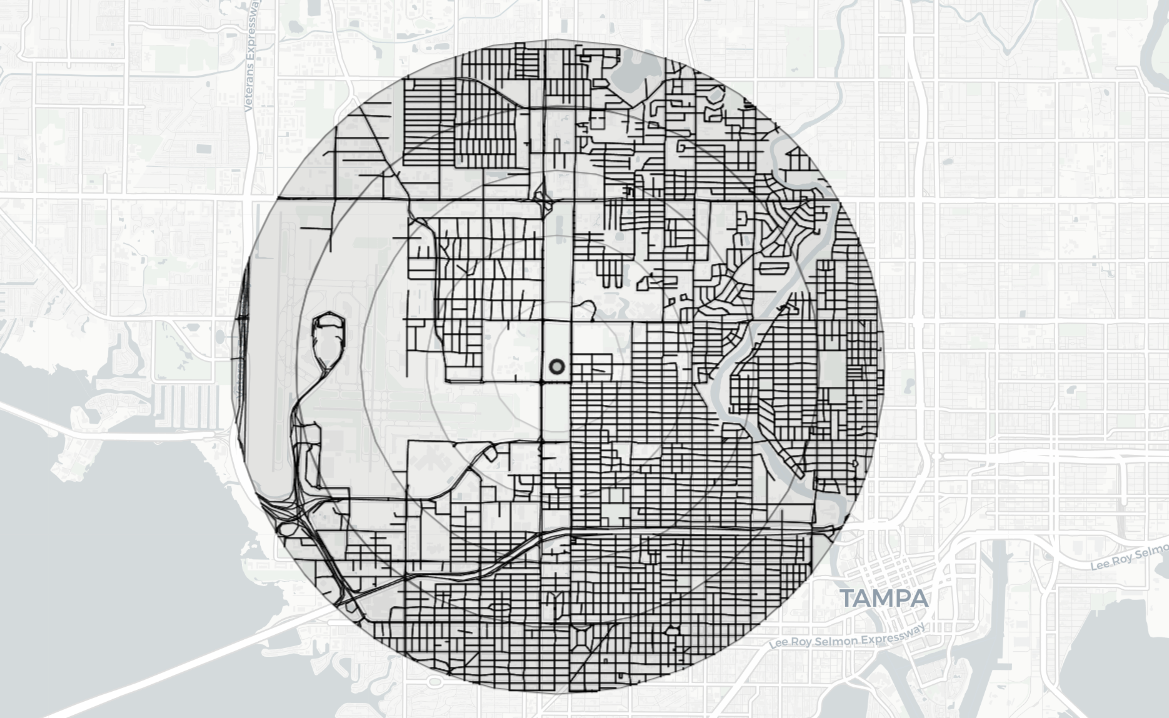}
\end{figure}

\begin{figure}[H]
\centering
\refstepcounter{figure}
\phantomsection
\label{fig:map_dolphins}
\captionsetup{justification=centering, font=small, labelfont=bf, labelsep=colon}
\caption*{\textbf{Figure  \thefigure}\\[4pt]0--5 km Radius Map for the Hard Rock Stadium (Miami Dolphins)}
\vspace{8pt}
\includegraphics[width=\textwidth]{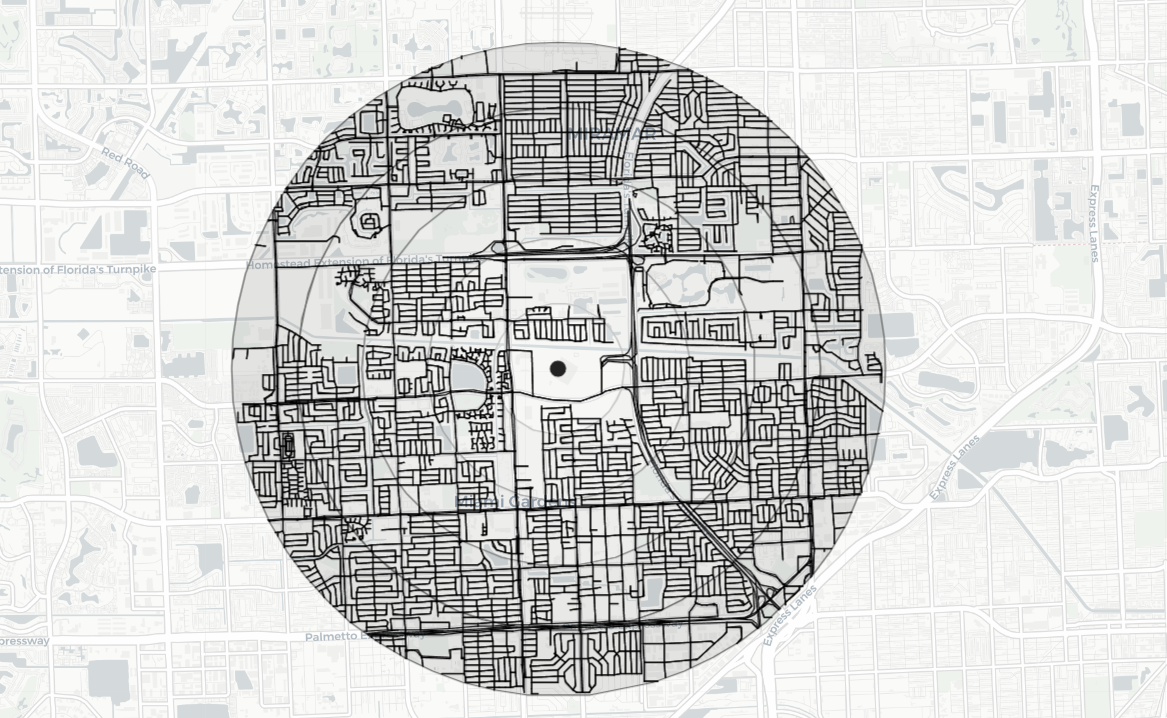}
\end{figure}

\begin{figure}[H]
\centering
\refstepcounter{figure}
\phantomsection
\label{fig:map_jags}
\captionsetup{justification=centering, font=small, labelfont=bf, labelsep=colon}
\caption*{\textbf{Figure  \thefigure}\\[4pt]0--5 km Radius Map for the EverBank Stadium (Jacksonville Jaguars)}
\vspace{8pt}
\includegraphics[width=\textwidth]{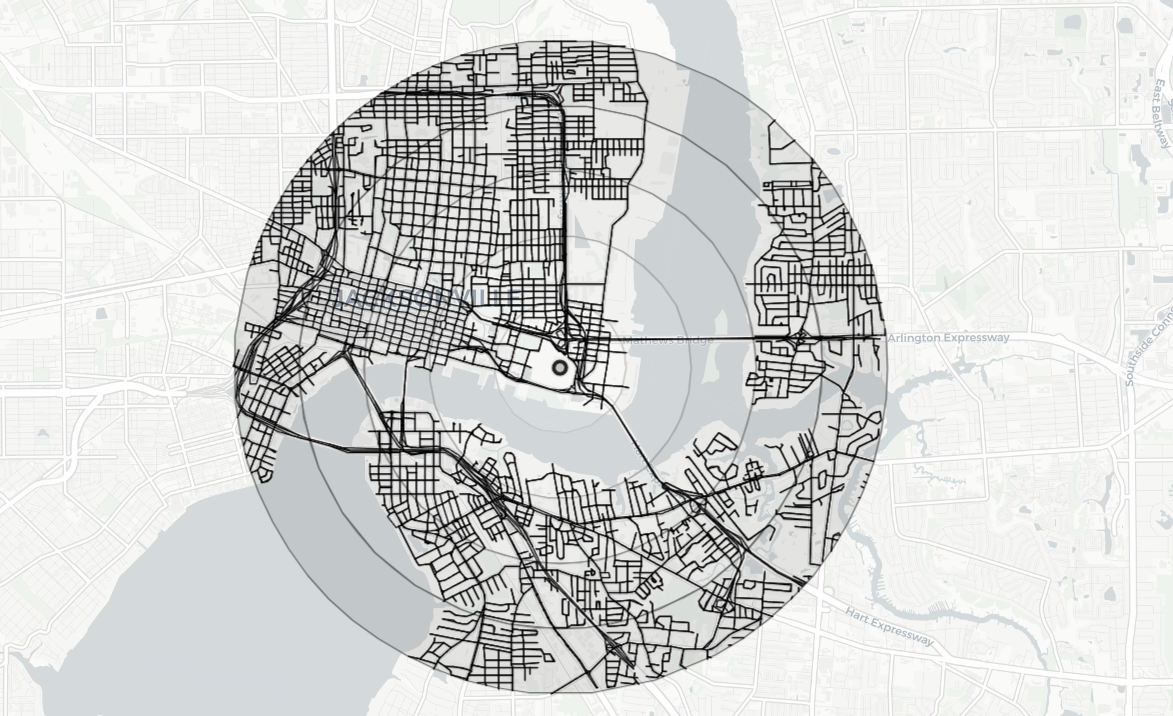}
\end{figure}

\begin{figure}[H]
\centering
\refstepcounter{figure}
\phantomsection
\label{fig:map_heat}
\captionsetup{justification=centering, font=small, labelfont=bf, labelsep=colon}
\caption*{\textbf{Figure  \thefigure}\\[4pt]0--5 km Radius Map for the Kaseya Center (Miami Heat)}
\vspace{8pt}
\includegraphics[width=\textwidth]{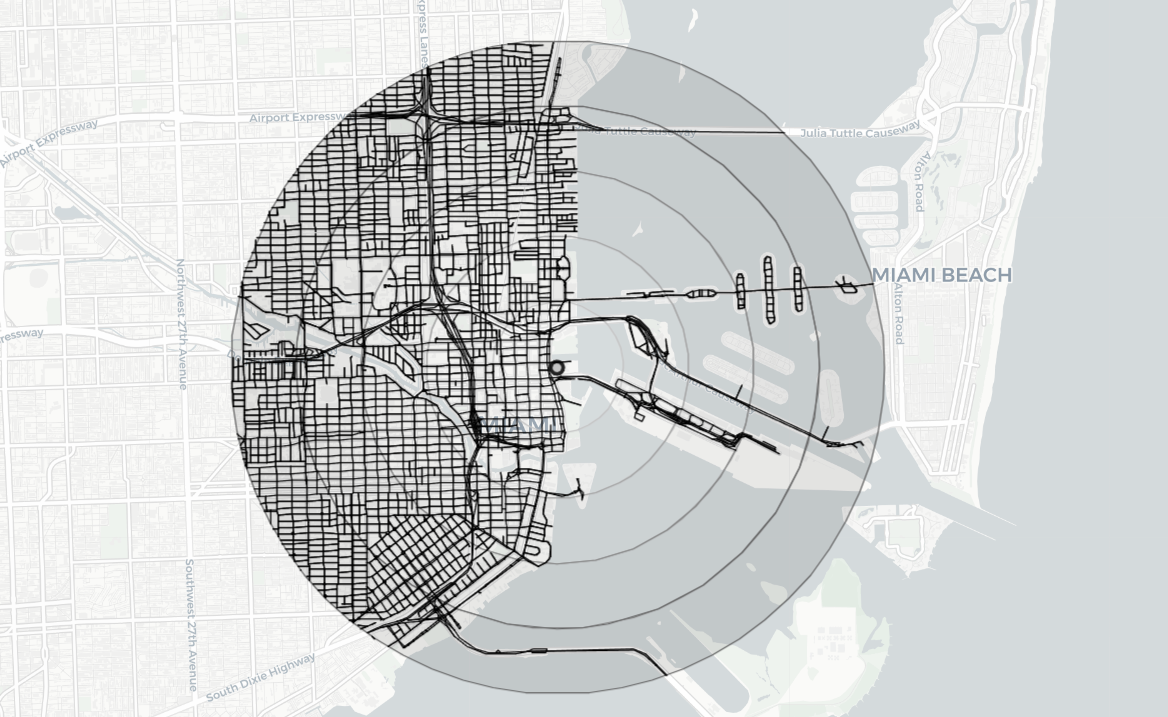}
\end{figure}

\begin{figure}[H]
\centering
\refstepcounter{figure}
\phantomsection
\label{fig:map_magic}
\captionsetup{justification=centering, font=small, labelfont=bf, labelsep=colon}
\caption*{\textbf{Figure  \thefigure}\\[4pt]0--5 km Radius Map for the Amway Center (Orlando Magic)}
\vspace{8pt}
\includegraphics[width=\textwidth]{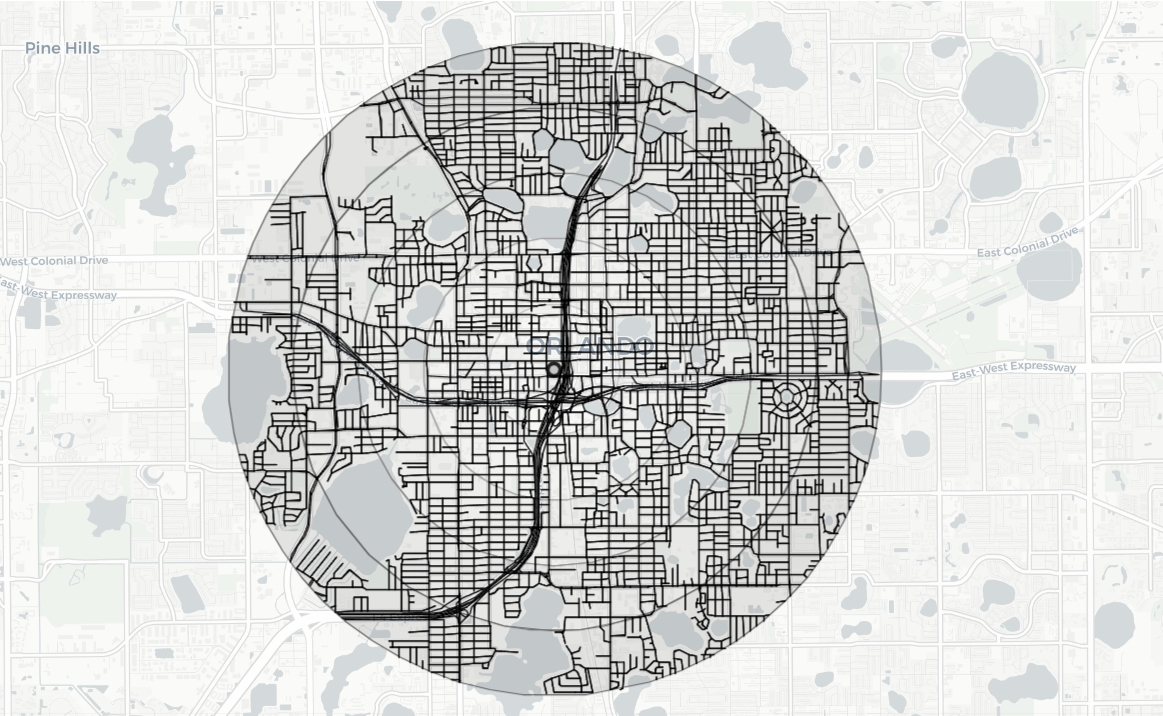}
\end{figure}


\newpage
\begin{figure}[H]
\centering
\refstepcounter{figure}
\label{fig:nfl_fans}
\captionsetup{justification=centering, font=small, labelfont=bf, labelsep=colon}
\caption*{\textbf{Figure  \thefigure}\\[4pt]Most-followed NFL teams per U.S. county}
\vspace{8pt}
\includegraphics[width=\textwidth]{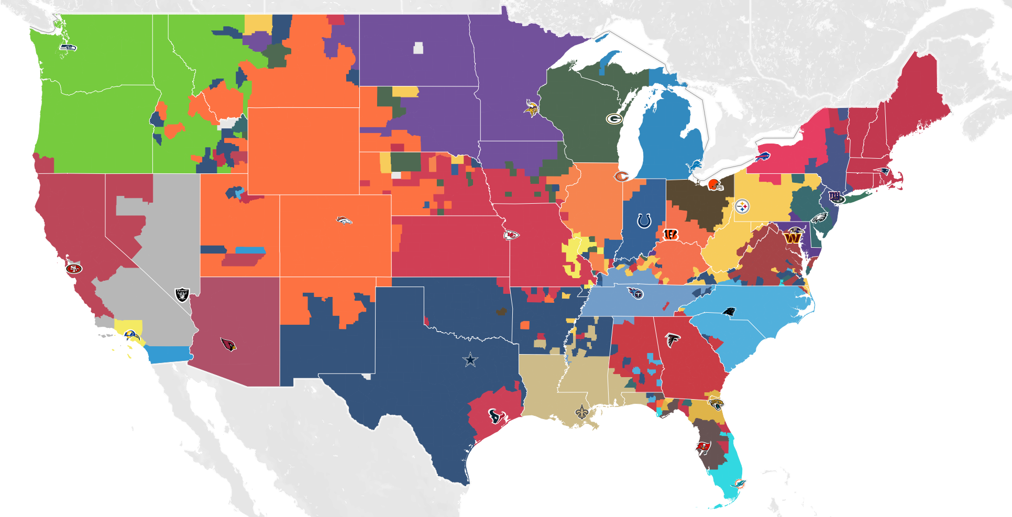}
\vspace{10pt}
\noindent
\begin{minipage}{\textwidth}
\begin{spacing}{0.9}
\scriptsize
\textit{Notes:} Source: \citet{SorensonNFL}.
\end{spacing}
\end{minipage}
\end{figure}

\begin{figure}[H]
\centering
\refstepcounter{figure}
\label{fig:florida_nfl_fans}
\captionsetup{justification=centering, font=small, labelfont=bf, labelsep=colon}
\caption*{\textbf{Figure  \thefigure}\\[4pt]Most-followed NFL teams by county in Florida}
\vspace{8pt}
\includegraphics[width=0.75\textwidth]{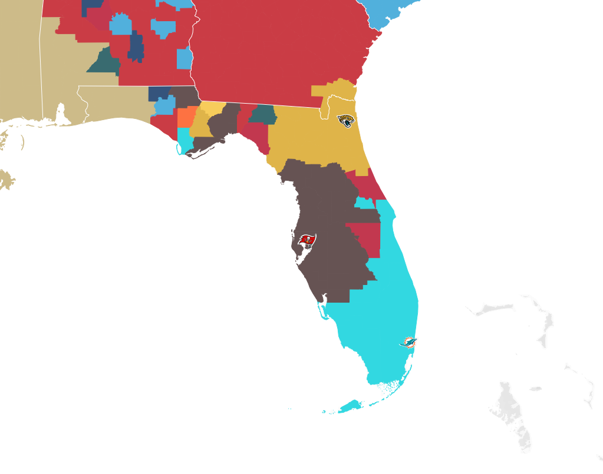}
\vspace{10pt}
\noindent
\begin{minipage}{\textwidth}
\begin{spacing}{0.9}
\scriptsize
\textit{Notes:} Source: \citet{SorensonNFL}. Brown: Tampa Bay Buccaneers; turquoise: Miami Dolphins; beige: Jacksonville Jaguars.
\end{spacing}
\end{minipage}
\end{figure}

\begin{figure}[H]
\centering
\refstepcounter{figure}
\label{fig:nba_fans}
\captionsetup{justification=centering, font=small, labelfont=bf, labelsep=colon}
\caption*{\textbf{Figure  \thefigure}\\[4pt]Most-followed NBA teams per U.S. county and Canadian census division}
\vspace{8pt}
\includegraphics[width=\textwidth]{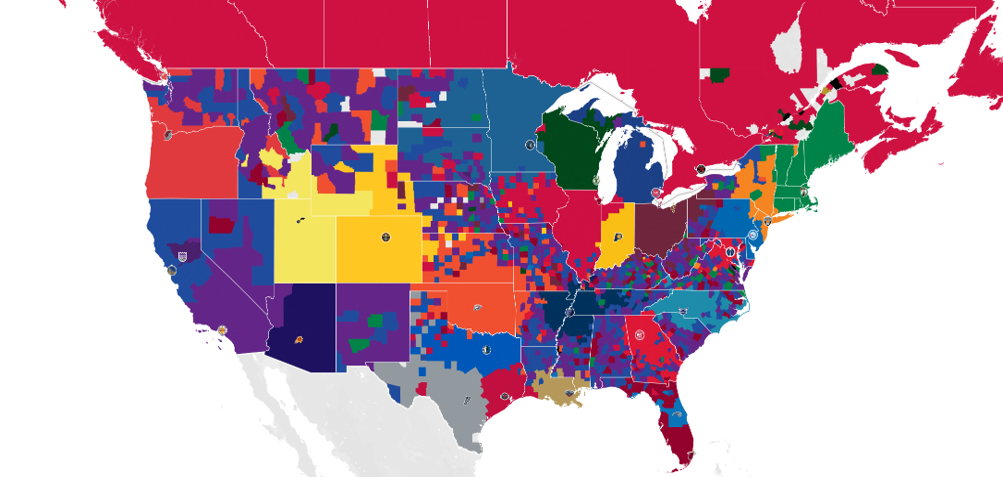}
\vspace{10pt}
\noindent
\begin{minipage}{\textwidth}
\begin{spacing}{0.9}
\scriptsize
\textit{Notes:} Source: \citet{SorensonNBA}.
\end{spacing}
\end{minipage}
\end{figure}

\begin{figure}[H]
\centering
\refstepcounter{figure}
\label{fig:florida_nba_fans}
\captionsetup{justification=centering, font=small, labelfont=bf, labelsep=colon}
\caption*{\textbf{Figure  \thefigure}\\[4pt]Most-followed NBA teams by county in Florida}
\vspace{8pt}
\includegraphics[width=0.75\textwidth]{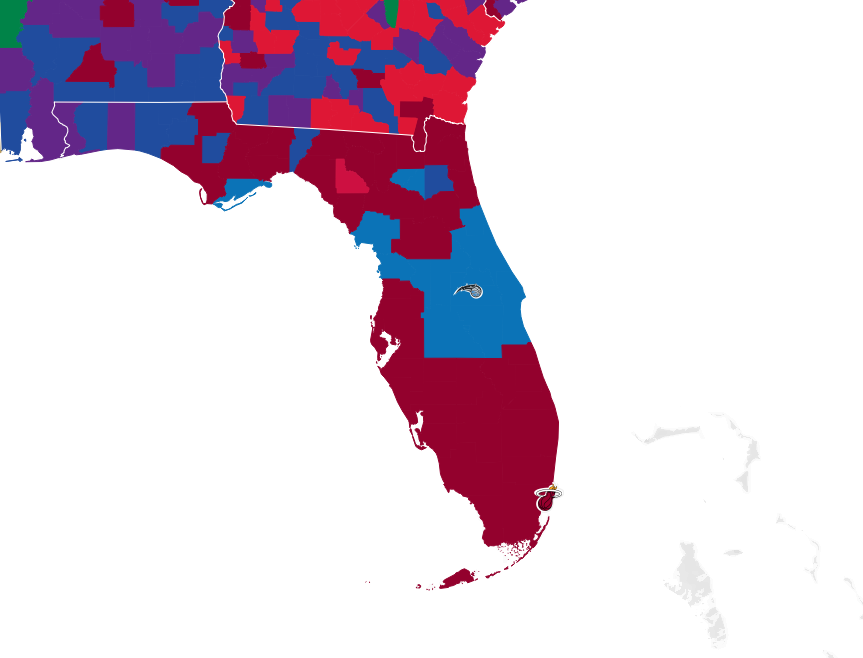}
\vspace{10pt}
\noindent
\begin{minipage}{\textwidth}
\begin{spacing}{0.9}
\scriptsize
\textit{Notes:} Source: \citet{SorensonNBA}. Dark red: Miami Heat; light blue: Orlando Magic.
\end{spacing}
\end{minipage}
\end{figure}


\begin{figure}[H]
\centering
\refstepcounter{figure}
\label{fig:nfl_twoway_regfe}
\captionsetup{justification=centering, font=small, labelfont=bf, labelsep=colon}
\caption*{\textbf{Figure \thefigure}\\[4pt]Effects of NFL game outcomes on vehicle speed near stadiums: two-way clustered standard errors}
\vspace{4pt}
\includegraphics[width=\textwidth]{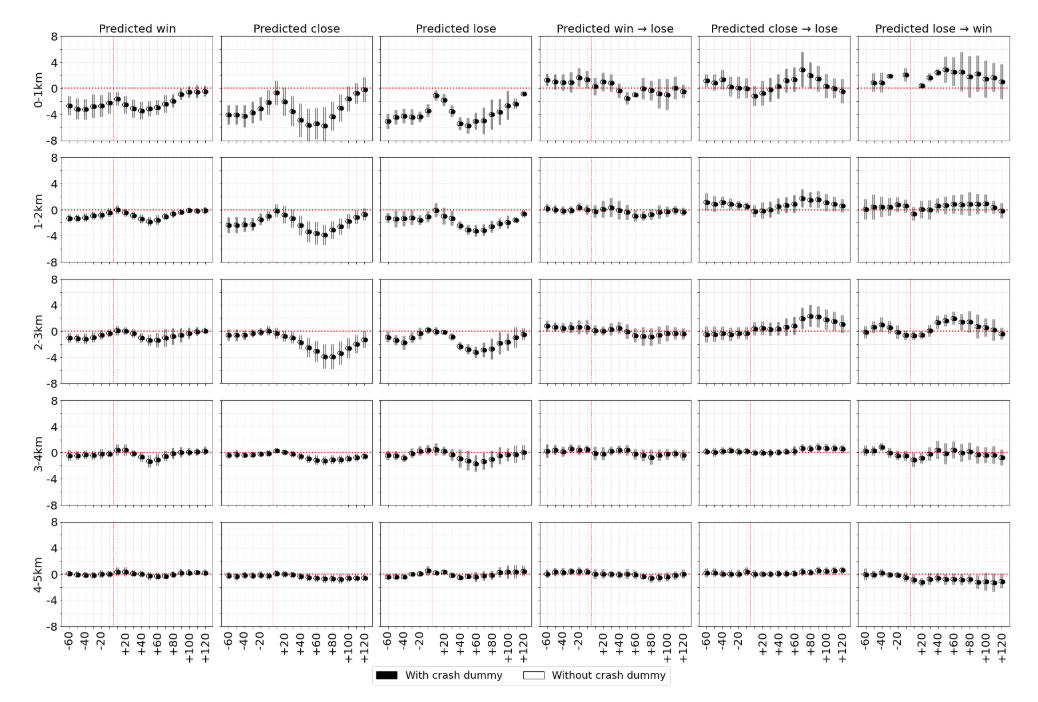}
\noindent
\begin{minipage}{\textwidth}
  \vspace{8pt}
  \begin{spacing}{0.9}
  \scriptsize
  \noindent\textit{Notes:} Predicted win indicates a point spread of --4 or less (negative spreads indicate the number of points a team is expected to win by); predicted close indicates a point spread between --4 and +4 exclusive; and predicted loss indicates a spread of +4 or more. This figure replicates the main specification using two-way clustered standard errors, clustering simultaneously at the TMC-code (entity) level and the season-week (time) level, following \citet{Cameron2011}. Fixed effects follow the baseline specification: week-of-season time fixed effects and season dummies. The sample is limited to Sundays during the regular-season home games for Florida's three NFL teams---the Miami Dolphins, Tampa Bay Buccaneers, and Jacksonville Jaguars---across the 2015--2019 seasons. Due to missing traffic data, the 2018 and 2019 Miami Dolphins seasons are excluded, resulting in a total of 84 Sunday home games with start times between 1:00 PM and 4:25 PM Eastern Time. Average vehicle speed represents observed speed across road links within 5 km of each stadium, broken up into 1-km distance bands, measured in 10-minute intervals spanning 60 minutes before to 120 minutes after each game.
  \end{spacing}
\end{minipage}
\end{figure}
\begin{figure}[H]
\centering
\refstepcounter{figure}
\label{fig:nfl_twoway_weekseasonfe}
\captionsetup{justification=centering, font=small, labelfont=bf, labelsep=colon}
\caption*{\textbf{Figure \thefigure}\\[4pt]Effects of NFL game outcomes on vehicle speed near stadiums: two-way clustered standard errors with season-week fixed effects}
\vspace{4pt}
\includegraphics[width=\textwidth]{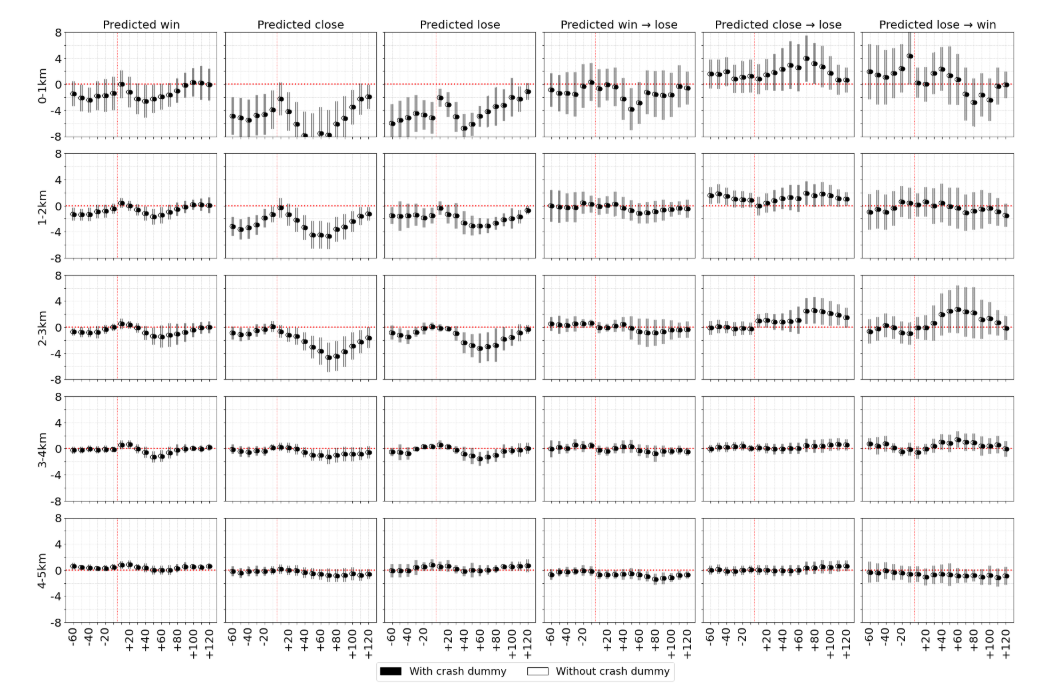}
\noindent
\begin{minipage}{\textwidth}
  \vspace{8pt}
  \begin{spacing}{0.9}
  \scriptsize
  \noindent\textit{Notes:} Predicted win indicates a point spread of --4 or less (negative spreads indicate the number of points a team is expected to win by); predicted close indicates a point spread between --4 and +4 exclusive; and predicted loss indicates a spread of +4 or more. This figure replicates the main specification using two-way clustered standard errors, clustering simultaneously at the TMC-code (entity) level and the season-week (time) level, following \citet{Cameron2011}. In contrast to Figure~\ref{fig:nfl_twoway_regfe}, fixed effects use a more demanding season-week specification comprising 84 unique time periods, with season dummies absorbed by the fixed effects. The sample is limited to Sundays during the regular-season home games for Florida's three NFL teams---the Miami Dolphins, Tampa Bay Buccaneers, and Jacksonville Jaguars---across the 2015--2019 seasons. Due to missing traffic data, the 2018 and 2019 Miami Dolphins seasons are excluded, resulting in a total of 84 Sunday home games with start times between 1:00 PM and 4:25 PM Eastern Time. Average vehicle speed represents observed speed across road links within 5 km of each stadium, broken up into 1-km distance bands, measured in 10-minute intervals spanning 60 minutes before to 120 minutes after each game.
  \end{spacing}
\end{minipage}
\end{figure}

\begin{figure}[H]
\centering
\refstepcounter{figure}
\label{fig:nfl_1pm}
\captionsetup{justification=centering, font=small, labelfont=bf, labelsep=colon}
\caption*{\textbf{Figure \thefigure}\\[4pt]Effects of NFL game outcomes on vehicle speed near stadiums: 1:00 PM-games sample}
\vspace{4pt}
\includegraphics[width=\textwidth]{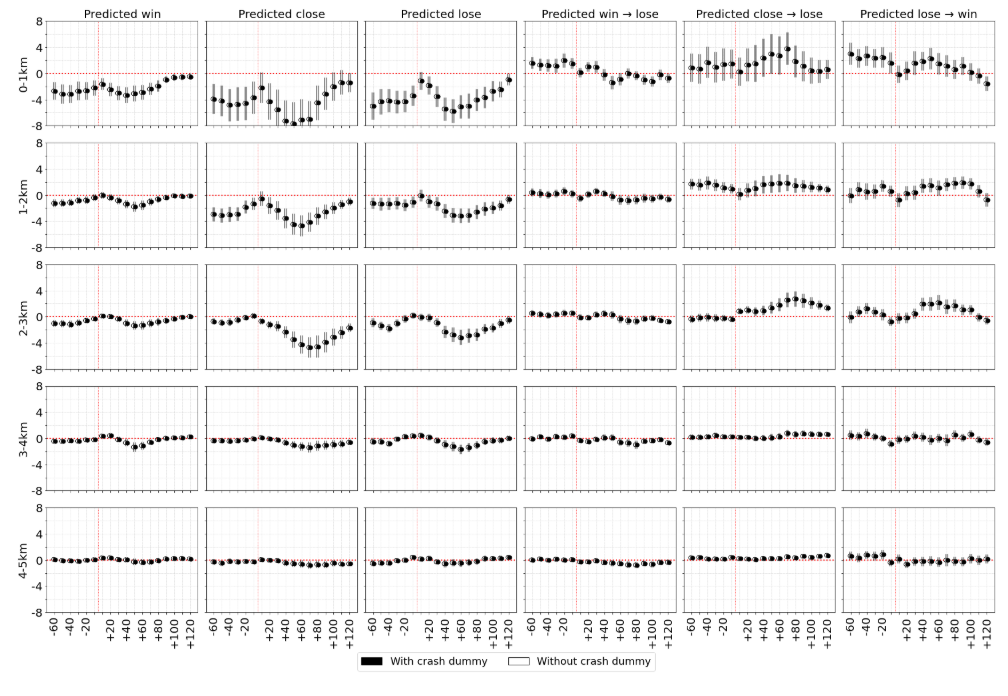}
\noindent
\begin{minipage}{\textwidth}
  \vspace{8pt}
  \begin{spacing}{0.9}
  \scriptsize
  \noindent\textit{Notes:} Predicted win indicates a point spread of --4 or less (negative spreads indicate the number of points a team is expected to win by); predicted close indicates a point spread between --4 and +4 exclusive; and predicted loss indicates a spread of +4 or more. The sample is restricted to 1:00 PM kickoff games only, and is otherwise limited to Sundays during the regular-season home games for Florida's three NFL teams---the Miami Dolphins, Tampa Bay Buccaneers, and Jacksonville Jaguars---across the 2015--2019 seasons. Due to missing traffic data, the 2018 and 2019 Miami Dolphins seasons are excluded, resulting in a total of 68 Sunday home games with a 1:00 PM start time. Regressions include TMC-code fixed effects, as well as week-of-season time fixed effects and season dummies to control for temporal variation. Standard errors are clustered at the TMC-code level. Average vehicle speed represents observed speed across road links within 5 km of each stadium, broken up into 1-km distance bands, measured in 10-minute intervals spanning 60 minutes before to 120 minutes after each game.
  \end{spacing}
\end{minipage}
\end{figure}

\begin{figure}[H]
\centering
\refstepcounter{figure}
\label{fig:nfl_nodolphins}
\captionsetup{justification=centering, font=small, labelfont=bf, labelsep=colon}
\caption*{\textbf{Figure \thefigure}\\[4pt]Effects of NFL game outcomes on vehicle speed near stadiums: excluding Miami Dolphins games}
\vspace{4pt}
\includegraphics[width=\textwidth]{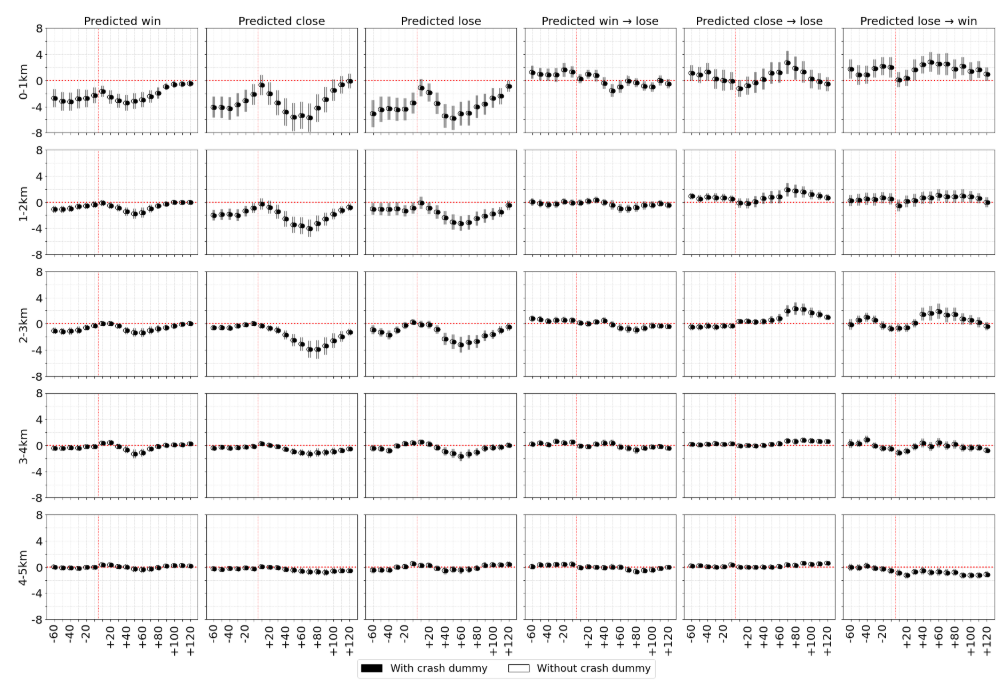}
\noindent
\begin{minipage}{\textwidth}
  \vspace{8pt}
  \begin{spacing}{0.9}
  \scriptsize
  \noindent\textit{Notes:} Predicted win indicates a point spread of --4 or less (negative spreads indicate the number of points a team is expected to win by); predicted close indicates a point spread between --4 and +4 exclusive; and predicted loss indicates a spread of +4 or more. The sample is limited to Sundays during the regular-season home games for Florida's three NFL teams across the 2015--2019 seasons, excluding all Miami Dolphins games due to missing traffic data in the Miami area. This results in 65 Sunday home games played by the Tampa Bay Buccaneers and Jacksonville Jaguars, with start times between 1:00 PM and 4:25 PM Eastern Time. Regressions include TMC-code fixed effects, as well as week-of-season time fixed effects and season dummies to control for temporal variation. Standard errors are clustered at the TMC-code level. Average vehicle speed represents observed speed across road links within 5 km of each stadium, broken up into 1-km distance bands, measured in 10-minute intervals spanning 60 minutes before to 120 minutes after each game.
  \end{spacing}
\end{minipage}
\end{figure}

\begin{figure}[H]
\centering
\refstepcounter{figure}
\label{fig:nfl_spread_3}
\captionsetup{justification=centering, font=small, labelfont=bf, labelsep=colon}
\caption*{\textbf{Figure \thefigure}\\[4pt]Effects of NFL game outcomes on vehicle speed near stadiums: $\pm$3 spread threshold}
\vspace{4pt}
\includegraphics[width=\textwidth]{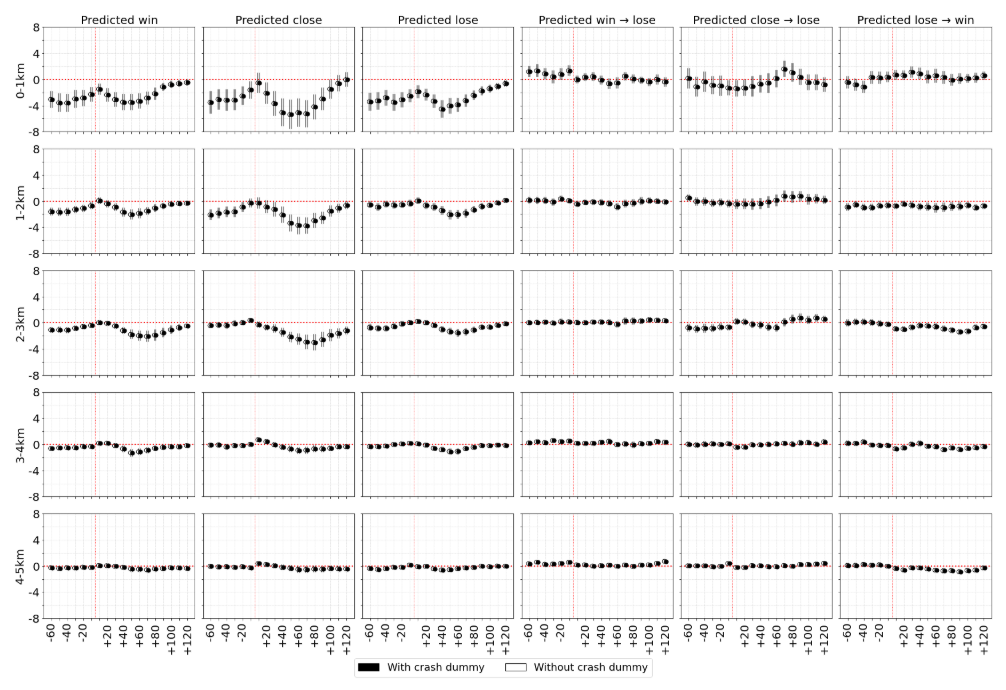}
\noindent
\begin{minipage}{\textwidth}
  \vspace{8pt}
  \begin{spacing}{0.9}
  \scriptsize
  \noindent\textit{Notes:} Predicted win indicates a point spread of --3 or less (negative spreads indicate the number of points a team is expected to win by); predicted close indicates a point spread between --3 and +3 exclusive; and predicted loss indicates a spread of +3 or more. This robustness check uses a narrower spread threshold of $\pm3$ points relative to the baseline specification of $\pm4$. The sample is limited to Sundays during the regular-season home games for Florida's three NFL teams---the Miami Dolphins, Tampa Bay Buccaneers, and Jacksonville Jaguars---across the 2015--2019 seasons. Due to missing traffic data, the 2018 and 2019 Miami Dolphins seasons are excluded, resulting in a total of 84 Sunday home games with start times between 1:00 PM and 4:25 PM Eastern Time. Regressions include TMC-code fixed effects, as well as week-of-season time fixed effects and season dummies to control for temporal variation. Standard errors are clustered at the TMC-code level. Average vehicle speed represents observed speed across road links within 5 km of each stadium, broken up into 1-km distance bands, measured in 10-minute intervals spanning 60 minutes before to 120 minutes after each game.
  \end{spacing}
\end{minipage}
\end{figure}

\begin{figure}[H]
\centering
\refstepcounter{figure}
\label{fig:nfl_spread_5}
\captionsetup{justification=centering, font=small, labelfont=bf, labelsep=colon}
\caption*{\textbf{Figure \thefigure}\\[4pt]Effects of NFL game outcomes on vehicle speed near stadiums: $\pm$5 spread threshold}
\vspace{4pt}
\includegraphics[width=\textwidth]{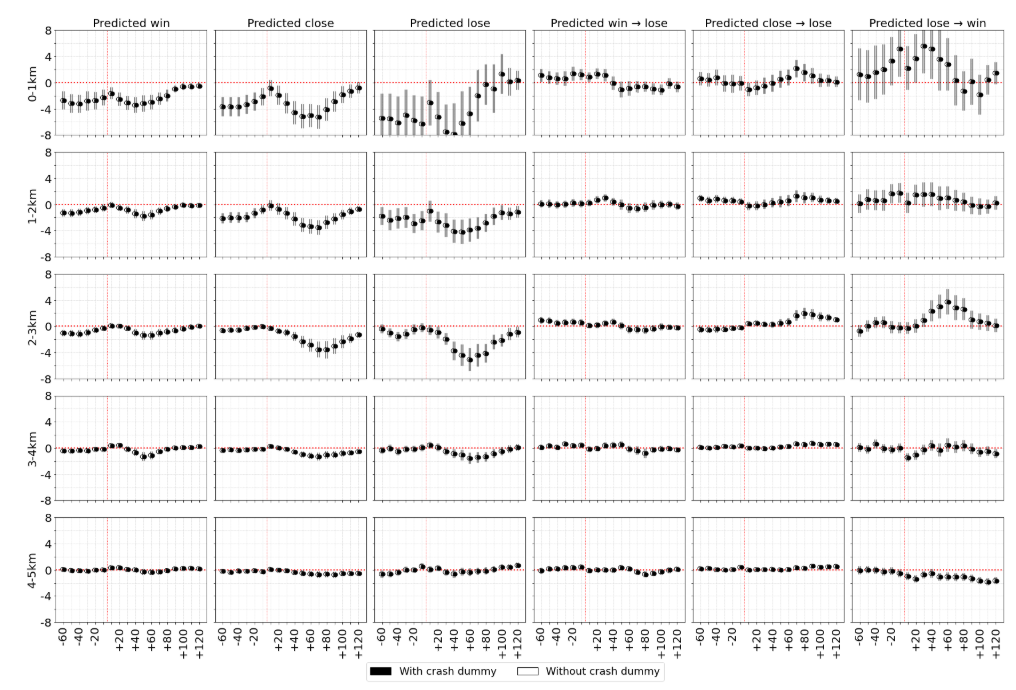}
\noindent
\begin{minipage}{\textwidth}
  \vspace{8pt}
  \begin{spacing}{0.9}
  \scriptsize
  \noindent\textit{Notes:} Predicted win indicates a point spread of --5 or less (negative spreads indicate the number of points a team is expected to win by); predicted close indicates a point spread between --5 and +5 exclusive; and predicted loss indicates a spread of +5 or more. This robustness check uses a wider spread threshold of $\pm5$ points relative to the baseline specification of $\pm4$. The sample is limited to Sundays during the regular-season home games for Florida's three NFL teams---the Miami Dolphins, Tampa Bay Buccaneers, and Jacksonville Jaguars---across the 2015--2019 seasons. Due to missing traffic data, the 2018 and 2019 Miami Dolphins seasons are excluded, resulting in a total of 84 Sunday home games with start times between 1:00 PM and 4:25 PM Eastern Time. Regressions include TMC-code fixed effects, as well as week-of-season time fixed effects and season dummies to control for temporal variation. Standard errors are clustered at the TMC-code level. Average vehicle speed represents observed speed across road links within 5 km of each stadium, broken up into 1-km distance bands, measured in 10-minute intervals spanning 60 minutes before to 120 minutes after each game.
  \end{spacing}
\end{minipage}
\end{figure}

\begin{figure}[H]
\centering
\refstepcounter{figure}
\label{fig:nfl_halftimecontrol}
\captionsetup{justification=centering, font=small, labelfont=bf, labelsep=colon}
\caption*{\textbf{Figure \thefigure}\\[4pt]Effects of NFL game outcomes on vehicle speed near stadiums: including halftime controls}
\vspace{4pt}
\includegraphics[width=\textwidth]{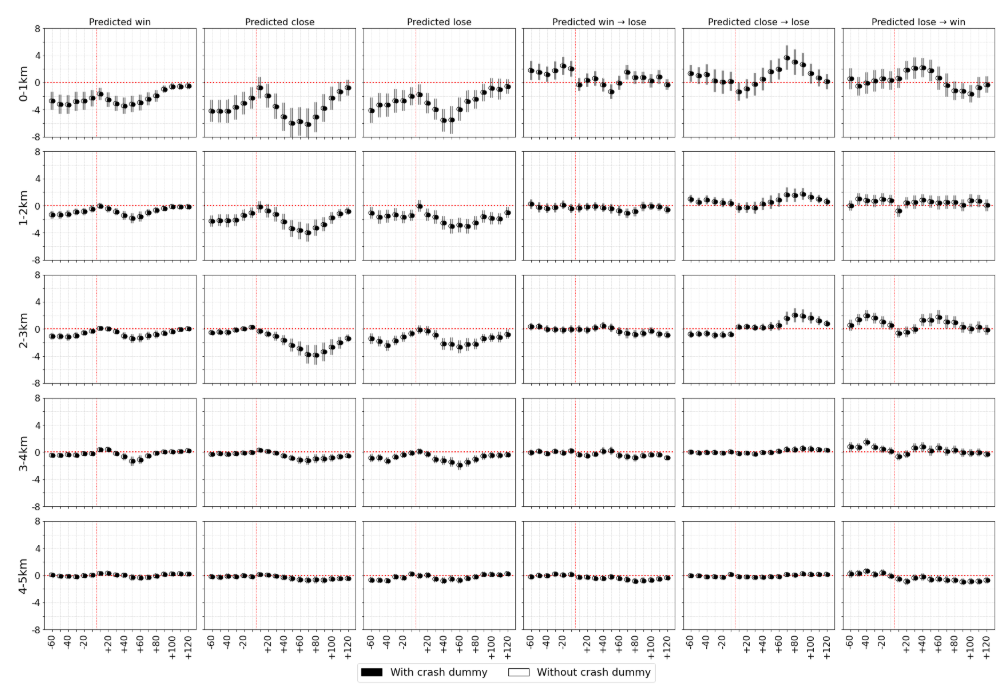}
\noindent
\begin{minipage}{\textwidth}
  \vspace{8pt}
  \begin{spacing}{0.9}
  \scriptsize
  \noindent\textit{Notes:} Predicted win indicates a point spread of --4 or less (negative spreads indicate the number of points a team is expected to win by); predicted close indicates a point spread between --4 and +4 exclusive; and predicted loss indicates a spread of +4 or more. In contrast to the baseline specification, this robustness check additionally controls for halftime status by including indicators for whether the home team was predicted to win, predicted close, or predicted to lose based on the halftime point differential, following \citet{Card2011}. The sample is limited to Sundays during the regular-season home games for Florida's three NFL teams---the Miami Dolphins, Tampa Bay Buccaneers, and Jacksonville Jaguars---across the 2015--2019 seasons. Due to missing traffic data, the 2018 and 2019 Miami Dolphins seasons are excluded, resulting in a total of 84 Sunday home games with start times between 1:00 PM and 4:25 PM Eastern Time. Regressions include TMC-code fixed effects, as well as week-of-season time fixed effects and season dummies to control for temporal variation. Standard errors are clustered at the TMC-code level. Average vehicle speed represents observed speed across road links within 5 km of each stadium, broken up into 1-km distance bands, measured in 10-minute intervals spanning 60 minutes before to 120 minutes after each game.
  \end{spacing}
\end{minipage}
\end{figure}

\begin{figure}[H]
\centering
\refstepcounter{figure}
\label{fig:nfl_halftimeupdate}
\captionsetup{justification=centering, font=small, labelfont=bf, labelsep=colon}
\caption*{\textbf{Figure \thefigure}\\[4pt]Effects of NFL game outcomes on vehicle speed near stadiums: halftime predictions}
\vspace{4pt}
\includegraphics[width=\textwidth]{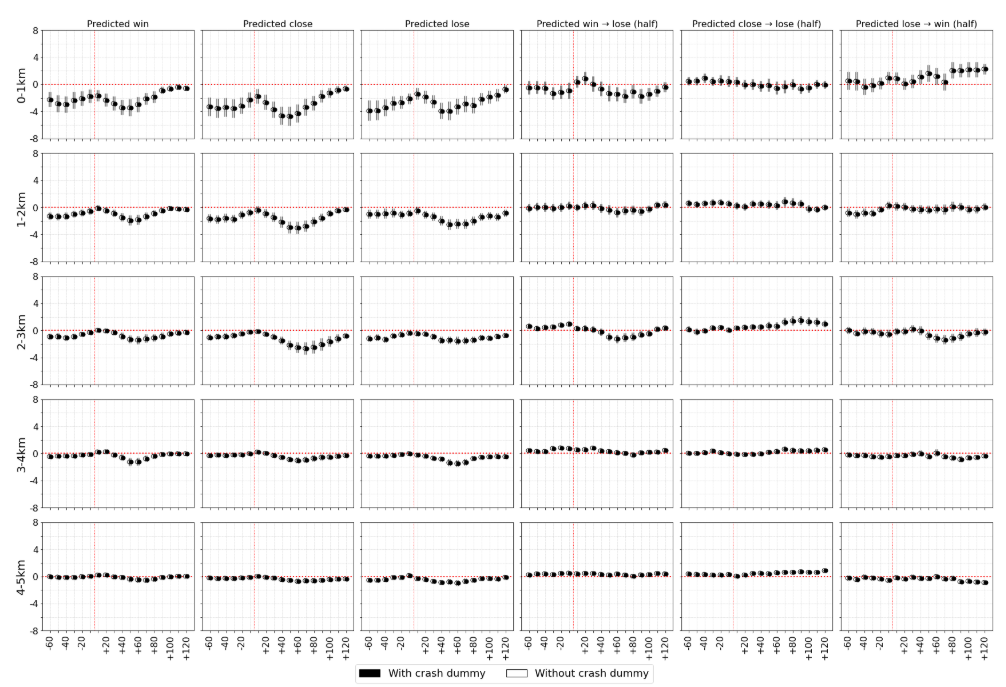}
\noindent
\begin{minipage}{\textwidth}
  \vspace{8pt}
  \begin{spacing}{0.9}
  \scriptsize
  \noindent\textit{Notes:} This figure shows how post-game traffic speed varies based on the halftime prediction rather than the pregame spread expectation, following \citet{Card2011}. Pregame predicted win, predicted close, and predicted loss are defined as in the baseline specification using a $\pm4$ point spread threshold. Halftime predicted win, halftime predicted close, and halftime predicted loss are defined based on the observed point differential at halftime, applying the same $\pm4$ threshold, where a negative halftime spread indicates the number of points the home team is actually leading by at halftime. The sample is limited to Sundays during the regular-season home games for Florida's three NFL teams---the Miami Dolphins, Tampa Bay Buccaneers, and Jacksonville Jaguars---across the 2015--2019 seasons. Due to missing traffic data, the 2018 and 2019 Miami Dolphins seasons are excluded, resulting in a total of 84 Sunday home games with start times between 1:00 PM and 4:25 PM Eastern Time. Regressions include TMC-code fixed effects, as well as week-of-season time fixed effects and season dummies to control for temporal variation. Standard errors are clustered at the TMC-code level. Average vehicle speed represents observed speed across road links within 5 km of each stadium, broken up into 1-km distance bands, measured in 10-minute intervals spanning 60 minutes before to 120 minutes after each game.
  \end{spacing}
\end{minipage}
\end{figure}

\FloatBarrier
\begin{figure}[H]
\centering
\refstepcounter{figure}
\label{fig:nfl_duringgame}
\captionsetup{justification=centering, font=small, labelfont=bf, labelsep=colon}
\caption*{\textbf{Figure \thefigure}\\[4pt]Traffic speed effects \textit{during} active NFL games}
\vspace{4pt}
\includegraphics[width=\textwidth]{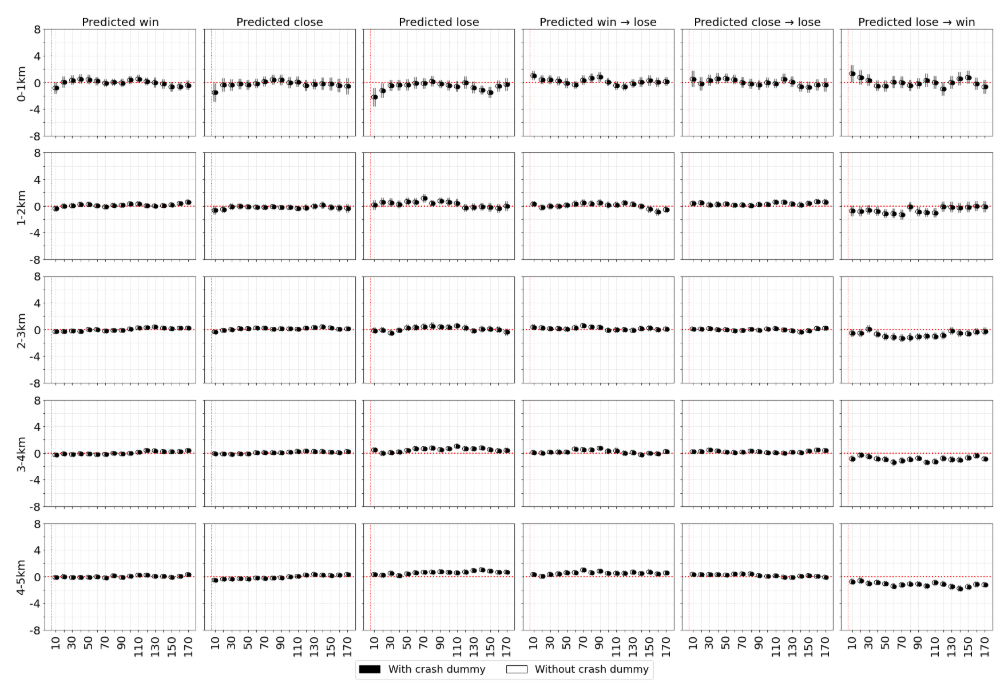}
\noindent
\begin{minipage}{\textwidth}
  \vspace{8pt}
  \begin{spacing}{0.9}
  \scriptsize
  \noindent\textit{Notes:} This figure shows how traffic speed varies during active NFL gameplay, comparing 10-minute intervals within games to matched non-gameday bins at the same time of day. Predicted win indicates a point spread of --4 or less (negative spreads indicate the number of points a team is expected to win by); predicted close indicates a point spread between --4 and +4 exclusive; and predicted loss indicates a spread of +4 or more. The sample is limited to Sundays during the regular-season home games for Florida's three NFL teams---the Miami Dolphins, Tampa Bay Buccaneers, and Jacksonville Jaguars---across the 2015--2019 seasons. Due to missing traffic data, the 2018 and 2019 Miami Dolphins seasons are excluded, resulting in a total of 84 Sunday home games with start times between 1:00 PM and 4:25 PM Eastern Time. Regressions include TMC-code fixed effects, as well as week-of-season time fixed effects and season dummies to control for temporal variation. Standard errors are clustered at the TMC-code level. Average vehicle speed represents observed speed across road links within 5 km of each stadium, broken up into 1-km distance bands, measured in 10-minute intervals spanning the first 170 minutes of gameplay---corresponding to the duration of the shortest game in the sample, ensuring all time intervals reflect strictly in-game periods and exclude any post-game traffic effects.
  \end{spacing}
\end{minipage}
\end{figure}
\FloatBarrier

\begin{figure}[H]
\centering
\refstepcounter{figure}
\label{fig:nfl_placebo_random}
\captionsetup{justification=centering, font=small, labelfont=bf, labelsep=colon}
\caption*{\textbf{Figure \thefigure}\\[4pt]Randomization placebo: average coefficients across 150 draws with randomly reassigned game outcomes}
\vspace{4pt}
\includegraphics[width=\textwidth]{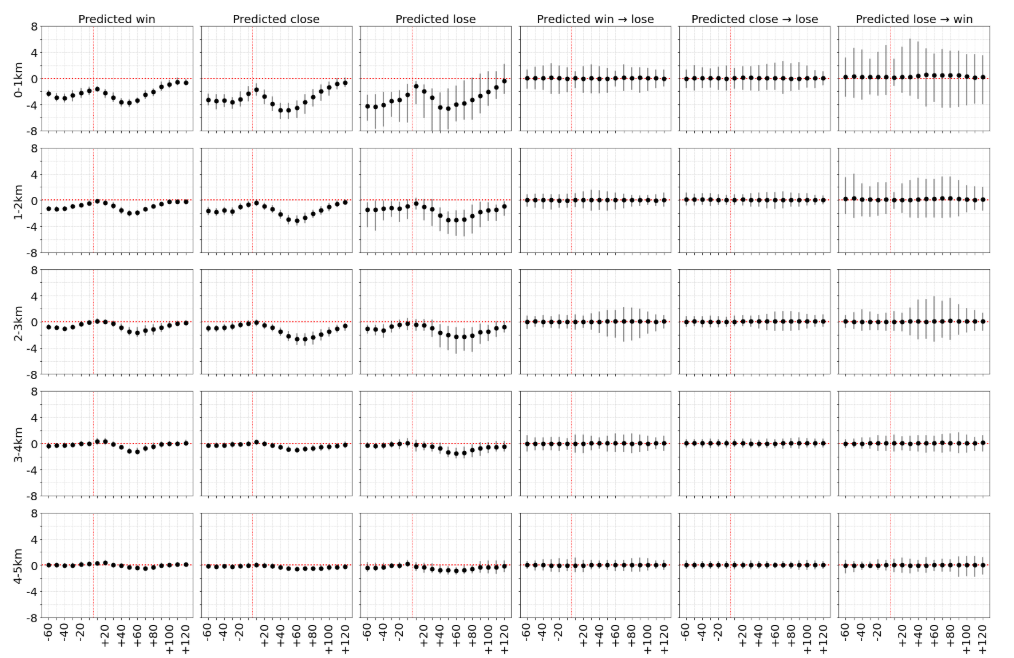}
\noindent
\begin{minipage}{\textwidth}
  \vspace{8pt}
  \begin{spacing}{0.9}
  \scriptsize
  \noindent\textit{Notes:} This figure presents the results of a randomization placebo test. The full regression model is run 150 times, each time randomly reassigning win/loss outcome labels within each spread category --- keeping spread category assignments fixed but randomizing the actual outcome within each category. All other aspects of the estimation remain identical to the baseline specification. Each subplot shows the average placebo coefficient across the 150 draws for each time window and distance band, with vertical lines representing the 2.5 and 97.5 percentile range of the 150 placebo coefficients, reflecting what would be expected by chance under random outcome assignment. Predicted win indicates a point spread of --4 or less; predicted close indicates a point spread between --4 and +4 exclusive; and predicted loss indicates a spread of +4 or more. The sample is limited to Sundays during the regular-season home games for Florida's three NFL teams---the Miami Dolphins, Tampa Bay Buccaneers, and Jacksonville Jaguars---across the 2015--2019 seasons. Due to missing traffic data, the 2018 and 2019 Miami Dolphins seasons are excluded, resulting in a total of 84 Sunday home games with start times between 1:00 PM and 4:25 PM Eastern Time. Regressions include TMC-code fixed effects, as well as week-of-season time fixed effects and season dummies to control for temporal variation. Standard errors are clustered at the TMC-code level. Average vehicle speed represents observed speed across road links within 5 km of each stadium, broken up into 1-km distance bands, measured in 10-minute intervals spanning 60 minutes before to 120 minutes after each game.
  \end{spacing}
\end{minipage}
\end{figure}

\begin{figure}[H]
\centering
\refstepcounter{figure}
\label{fig:nfl_placebo_pseudogame}
\captionsetup{justification=centering, font=small, labelfont=bf, labelsep=colon}
\caption*{\textbf{Figure \thefigure}\\[4pt]Pseudo game-day placebo: average coefficients across 150 draws with real game labels assigned to non-game Sundays}
\vspace{4pt}
\includegraphics[width=\textwidth]{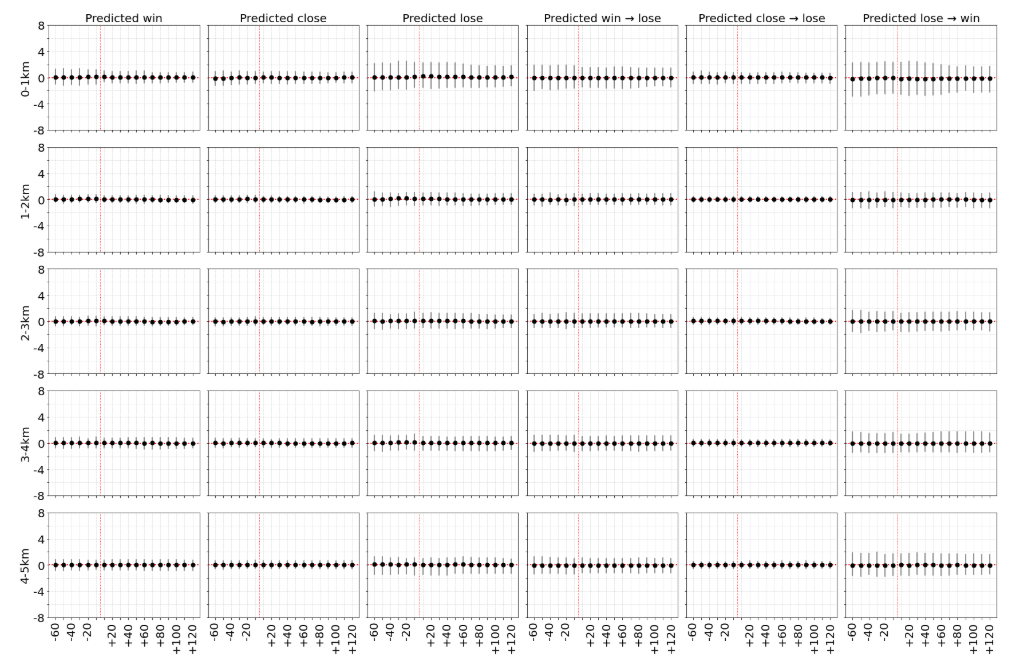}
\noindent
\begin{minipage}{\textwidth}
  \vspace{8pt}
  \begin{spacing}{0.9}
  \scriptsize
  \noindent\textit{Notes:} This figure presents the results of a pseudo game-day placebo test. The sample is restricted to non-game Sundays only. Each of the 84 real games is randomly assigned to a non-game Sunday from the same team's stadium area, without replacement, and the assigned date inherits the real game's spread category and outcome label. The existing matched time-window indicators, which flag observations occurring at the same time of day as a real game's pre- and post-game windows, are used as event-time dummies for the pseudo game days. The remaining non-game Sundays serve as the control group. This procedure is repeated 150 times, each time with a new random assignment of real games to non-game Sundays. The figure presents the average coefficient across the 150 draws for each time window and distance band, along with the 2.5th and 97.5th percentile range of the placebo coefficients. The same regression specification as the main model is estimated on this pseudo game-day dataset using only non-game Sunday speed data throughout. Predicted win indicates a point spread of --4 or less; predicted close indicates a point spread between --4 and +4 exclusive; and predicted loss indicates a spread of +4 or more. The sample is limited to Sundays during the regular-season home games for Florida's three NFL teams---the Miami Dolphins, Tampa Bay Buccaneers, and Jacksonville Jaguars---across the 2015--2019 seasons. Due to missing traffic data, the 2018 and 2019 Miami Dolphins seasons are excluded, resulting in a total of 84 Sunday home games with start times between 1:00 PM and 4:25 PM Eastern Time. Regressions include TMC-code fixed effects, as well as week-of-season time fixed effects and season dummies to control for temporal variation. Standard errors are clustered at the TMC-code level. Average vehicle speed represents observed speed across road links within 5 km of each stadium, broken up into 1-km distance bands, measured in 10-minute intervals spanning 60 minutes before to 120 minutes after each game.
  \end{spacing}
\end{minipage}
\end{figure}

\clearpage

\begin{table}[H]
\centering
\refstepcounter{table}
\phantomsection
\label{tab:nfl_crash}
\captionsetup{justification=centering, font=small, labelfont=bf, labelsep=colon}

\caption*{\textbf{Table \thetable}\\[4pt]Poisson fixed-effects regressions of NFL crash counts by predicted outcome and actual result}
\vspace{6pt}
\scriptsize
\setlength{\tabcolsep}{0pt}
\renewcommand{\arraystretch}{0.7}
\begin{tabular*}{\textwidth}{@{\extracolsep{\fill}}lccccc@{}}
\toprule
\textbf{Parameter} & \makecell{\textbf{Number of crashes}} & \textbf{Estimate} & \textbf{$p$-value} & \textbf{95\% CI Low} & \textbf{95\% CI High} \\
\midrule
\textbf{Pregame\_60} & & & & & \\
Gameday & 17 & 0.933 & 0.008 &  0.249 & 1.616 \\
Predicted win & 4 &  0.417 & 0.545 & -0.934 & 1.768 \\
Predicted close & 9 &  0.169 & 0.826 & -1.341 & 1.679 \\
Predicted lose & 4 &  2.337 & 0.010 &  0.551 & 4.123 \\
Predicted win $\rightarrow$ lose & 1 &  0.528 & 0.680 & -1.978 & 3.034 \\
Predicted close $\rightarrow$ lose & 7 &  0.920 & 0.281 & -0.752 & 2.593 \\
Predicted lose $\rightarrow$ win & 1 & -0.810 & 0.581 & -3.689 & 2.069 \\
\midrule
\textbf{Postgame\_60} & & & & & \\
Gameday & 40 & 1.618 & 0.000 &  1.159 & 2.077 \\
Predicted win & 13 &  1.480 & 0.001 &  0.613 & 2.348 \\
Predicted close & 25 &  1.745 & 0.000 &  0.884 & 2.606 \\
Predicted lose & 2 &  0.927 & 0.458 & -1.522 & 3.376 \\
Predicted win $\rightarrow$ lose & 8 &  0.951 & 0.138 & -0.305 & 2.207 \\
Predicted close $\rightarrow$ lose & 13 & -0.248 & 0.565 & -1.091 & 0.596 \\
Predicted lose $\rightarrow$ win & 1 & -0.353 & 0.831 & -3.590 & 2.883 \\
\midrule
\textbf{Postgame\_120} & & & & & \\
Gameday & 21 & 1.232 & 0.000 &  0.662 & 1.803 \\
Predicted win & 5 &  1.452 & 0.040 &  0.066 & 2.838 \\
Predicted close & 13 &  0.467 & 0.458 & -0.766 & 1.700 \\
Predicted lose & 3 &  0.879 & 0.420 & -1.256 & 3.015 \\
Predicted win $\rightarrow$ lose & 1 & -0.663 & 0.590 & -3.075 & 1.749 \\
Predicted close $\rightarrow$ lose & 10 &  0.967 & 0.173 & -0.424 & 2.358 \\
Predicted lose $\rightarrow$ win & 2 &  1.003 & 0.479 & -1.776 & 3.783 \\
\bottomrule
\end{tabular*}
\noindent
\begin{minipage}{\textwidth}
  \vspace{10pt}
  \begin{spacing}{0.9}
  \scriptsize
  \noindent\textit{Notes:} Poisson fixed-effects regressions of crash counts on predicted outcome categories and interaction terms, estimated on the aggregated 0--5 km sample. Crash counts reported in this table reflect the total number of unique crash events aggregated across all observation windows in the 2015--2019 sample (i.e., all gamedays and all corresponding matched non-gameday control periods), rather than averages per gameday or per non-gameday. Predicted win indicates a point spread of --4 or less (negative spreads indicate the number of points a team is expected to win by); predicted close indicates a point spread between --4 and +4 exclusive; and predicted loss indicates a spread of +4 or more. The sample is limited to Sundays during the regular-season home games for Florida's three NFL teams---the Miami Dolphins, Tampa Bay Buccaneers, and Jacksonville Jaguars---across the 2015--2019 seasons. Due to missing traffic data, the 2018 and 2019 Miami Dolphins seasons are excluded, resulting in a total of 84 Sunday home games with start times between 1:00 PM and 4:25 PM Eastern Time. Poisson regressions include TMC-code fixed effects, as well as week-of-season time fixed effects and season dummies to control for temporal variation. Standard errors are clustered at the TMC-code level. Coefficients are reported as log changes; positive values indicate higher crash frequencies relative to matched non-gameday periods (or alternative outcomes).
  \end{spacing}
\end{minipage}
\end{table}

\begin{table}[H]
\centering
\refstepcounter{table}
\phantomsection
\label{tab:nba_crash}
\captionsetup{justification=centering, font=small, labelfont=bf, labelsep=colon}

\caption*{\textbf{Table \thetable}\\[4pt]Poisson fixed-effects regressions of NBA crash counts by predicted outcome and actual result}
\vspace{6pt}
\scriptsize
\setlength{\tabcolsep}{0pt}
\renewcommand{\arraystretch}{0.7}
\begin{tabular*}{\textwidth}{@{\extracolsep{\fill}}lccccc@{}}
\toprule
\textbf{Parameter} & \makecell{\textbf{Number of crashes}} & \textbf{Estimate} & \textbf{$p$-value} & \textbf{95\% CI Low} & \textbf{95\% CI High} \\
\midrule
\textbf{Pregame\_60} & & & & & \\
Gameday & 89 &  0.003 & 0.978 & -0.240 &  0.247 \\
Predicted win & 9 & -0.309 & 0.406 & -1.039 &  0.421 \\
Predicted close & 60 &  0.217 & 0.189 & -0.107 &  0.540 \\
Predicted lose & 20 & -0.116 & 0.727 & -0.764 &  0.533 \\
Predicted win $\rightarrow$ lose & 2 & -0.491 & 0.648 & -2.601 &  1.618 \\
Predicted close $\rightarrow$ lose & 26 & -0.308 & 0.148 & -0.726 &  0.109 \\
Predicted lose $\rightarrow$ win & 9 &  0.439 & 0.383 & -0.548 &  1.426 \\
\midrule
\textbf{Postgame\_60} & & & & & \\
Gameday & 40 &  0.233 & 0.228 & -0.145 &  0.611 \\
Predicted win & 5 &  0.185 & 0.693 & -0.733 &  1.103 \\
Predicted close & 31 &  0.340 & 0.280 & -0.277 &  0.958 \\
Predicted lose & 4 & -1.641 & 0.103 & -3.612 &  0.330 \\
Predicted win $\rightarrow$ lose & 0 & \multicolumn{4}{c}{No crash occurred} \\
Predicted close $\rightarrow$ lose & 18 &  0.313 & 0.408 & -0.427 &  1.052 \\
Predicted lose $\rightarrow$ win & 3 &  1.850 & 0.141 & -0.613 &  4.314 \\
\midrule
\textbf{Postgame\_120} & & & & & \\
Gameday & 29 & -0.095 & 0.703 & -0.584 &  0.394 \\
Predicted win & 7 &  0.072 & 0.876 & -0.830 &  0.974 \\
Predicted close & 17 & -0.113 & 0.765 & -0.852 &  0.627 \\
Predicted lose & 5 & -0.506 & 0.369 & -1.609 &  0.598 \\
Predicted win $\rightarrow$ lose & 2 &  0.577 & 0.503 & -1.112 &  2.266 \\
Predicted close $\rightarrow$ lose & 9 &  0.040 & 0.931 & -0.858 &  0.937 \\
Predicted lose $\rightarrow$ win & 2 &  0.261 & 0.778 & -1.555 &  2.076 \\
\bottomrule
\end{tabular*}
\noindent
\begin{minipage}{\textwidth}
  \vspace{10pt}
  \begin{spacing}{0.9}
  \scriptsize
  \noindent\textit{Notes:} Poisson fixed-effects regressions of crash counts on predicted outcome categories and interaction terms, estimated on the aggregated 0--5 km sample. Crash counts reported in this table reflect the total number of unique crash events aggregated across all observation windows in the 2015/16--2018/19 sample (i.e., all gamedays and all corresponding matched non-gameday control periods), rather than averages per gameday or per non-gameday. Predicted win indicates a point spread of --6.75 or less (negative spreads indicate the number of points a team is expected to win by); predicted close indicates a point spread between --6.75 and +4.75 exclusive; and predicted loss indicates a spread of +4.75 or more. The sample is limited to regular-season NBA games between the 2015/16 and 2018/19 seasons, encompassing four total seasons. The analysis includes home games played by Florida's two NBA teams---the Miami Heat and the Orlando Magic---with scheduled tip-off times of 6:00 PM or later. Regressions include TMC-code fixed effects, as well as month time fixed effects, season dummies, and day-of-week dummies to control for temporal variation. Standard errors are clustered at the TMC-code level. Coefficients are reported as log changes; positive values indicate higher crash frequencies relative to matched non-gameday periods (or alternative outcomes).
  \end{spacing}
\end{minipage}
\end{table}

\clearpage
\begin{table}[H]
\centering
\refstepcounter{table}
\phantomsection
\label{tab:nfl_table}
\captionsetup{justification=centering, font=small, labelfont=bf, labelsep=colon}
\caption*{\textbf{Table \thetable}\\[4pt]NFL Regression results by time and distance}
\end{table}
\vspace{-0.75\baselineskip}
\begingroup
\setstretch{1}
\scriptsize
\setlength{\tabcolsep}{0pt}
\renewcommand{\arraystretch}{0.7}

\setlength{\LTpre}{0pt}
\setlength{\LTpost}{0pt}
\setlength{\LTleft}{0pt}
\setlength{\LTright}{0pt}

\begin{longtable}{@{\extracolsep{\fill}}lllrrrr@{}}
\\
\toprule
\textbf{Time} & 
\textbf{Distance} & 
\textbf{Variable} & 
\textbf{With Crash} & 
\textbf{With Crash SE} & 
\textbf{No Crash} & 
\textbf{No Crash SE} \\
\midrule
\endfirsthead
\toprule
\textbf{Time} & 
\textbf{Distance} & 
\textbf{Variable} & 
\textbf{With Crash} & 
\textbf{With Crash SE} & 
\textbf{No Crash} & 
\textbf{No Crash SE} \\
\midrule
\endhead

\midrule
\multicolumn{7}{r}{\emph{Continued on next page}} \\
\endfoot

\bottomrule
\endlastfoot

\multicolumn{7}{l}{\textbf{Pregame 60}}\\
\addlinespace[2pt]
 
& \multicolumn{1}{l}{\hspace{1.5em}0--1 km} & & & & & \\
& & Predicted Win             &  -2.7250 &  0.6613 &  -2.7250 &  0.6613 \\
& & Predicted Close           &  -4.1019 &  0.7760 &  -4.1024 &  0.7760 \\
& & Predicted Lose            &  -5.0876 &  1.0321 &  -5.0875 &  1.0321 \\
& & Predicted Win Lose        &   1.2369 &  0.4517 &   1.2368 &  0.4517 \\
& & Predicted Close Lose      &   1.1157 &  0.5589 &   1.1162 &  0.5590 \\
& & Predicted Lose Win        &   1.7266 &  0.7203 &   1.7263 &  0.7203 \\
\end{longtable}
\noindent
\begin{minipage}{\textwidth}
  \vspace{10pt}
  \begin{spacing}{0.9}
  \scriptsize
  \noindent\textit{Notes:} This table reports the coefficients corresponding to Figure~\ref{fig:nfl}. For brevity, coefficients for crash, weather, holiday, and season dummy indicators are omitted; this amounts to 1{,}980 omitted coefficients across the 180 regressions reported in the table. Predicted win indicates a point spread of --4 or less (negative spreads indicate the number of points a team is expected to win by); predicted close indicates a point spread between --4 and +4 exclusive; and predicted loss indicates a spread of +4 or more. The sample is limited to Sundays during the regular-season home games for Florida's three NFL teams---the Miami Dolphins, Tampa Bay Buccaneers, and Jacksonville Jaguars---across the 2015--2019 seasons. Due to missing traffic data, the 2018 and 2019 Miami Dolphins seasons are excluded, resulting in a total of 84 Sunday home games with start times between 1:00 PM and 4:25 PM Eastern Time. Regressions include TMC-code fixed effects, as well as week-of-season time fixed effects and season dummies to control for temporal variation. Standard errors are clustered at the TMC-code level. Average vehicle speed represents observed speed across road links within 5 km of each stadium, broken up into 1-km distance bands, measured in 10-minute intervals from 60 minutes before to 120 minutes after each game.
  \end{spacing}
\end{minipage}
\endgroup

\addtocounter{table}{-1}
\clearpage
\begin{table}[H]
\centering
\refstepcounter{table}
\phantomsection
\label{tab:nba_table}
\captionsetup{justification=centering, font=small, labelfont=bf, labelsep=colon}
\caption*{\textbf{Table \thetable}\\[4pt]NBA Regression results by time and distance}
\end{table}
\vspace{-0.75\baselineskip}
\begingroup
\setstretch{1}
\scriptsize
\setlength{\tabcolsep}{0pt}
\renewcommand{\arraystretch}{0.7}
\setlength{\LTpre}{0pt}
\setlength{\LTpost}{0pt}
\setlength{\LTleft}{0pt}
\setlength{\LTright}{0pt}

\begin{longtable}{@{\extracolsep{\fill}}lllrrrr@{}}
\label{tab:nba_regression}\\
\toprule
\textbf{Time} & \textbf{Distance} & \textbf{Variable} & 
\textbf{With Crash} & \textbf{With Crash SE} & 
\textbf{No Crash} & \textbf{No Crash SE} \\
\midrule
\endfirsthead
\toprule
\textbf{Time} & \textbf{Distance} & \textbf{Variable} & 
\textbf{With Crash} & \textbf{With Crash SE} & 
\textbf{No Crash} & \textbf{No Crash SE} \\
\midrule
\endhead
\midrule
\multicolumn{7}{r}{\emph{Continued on next page}} \\
\endfoot
\bottomrule
\endlastfoot

\multicolumn{7}{l}{\textbf{Pregame 60}}\\
\addlinespace[2pt]

& \multicolumn{1}{l}{\hspace{1.5em}0--1 km} & & & & & \\
& & Predicted Win             &   0.2638 &  0.1256 &   0.2641 &  0.1256 \\
& & Predicted Close           &   0.6235 &  0.1829 &   0.6232 &  0.1829 \\
& & Predicted Lose            &  -0.0653 &  0.0833 &  -0.0646 &  0.0835 \\
& & Predicted Win Lose        &   0.0170 &  0.2218 &   0.0173 &  0.2219 \\
& & Predicted Close Lose      &  -0.4269 &  0.0784 &  -0.4273 &  0.0786 \\
& & Predicted Lose Win        &   0.1018 &  0.1148 &   0.1017 &  0.1148 \\
\end{longtable}
\noindent
\begin{minipage}{\textwidth}
  \vspace{10pt}
  \begin{spacing}{0.9}
  \scriptsize
  \noindent\textit{Notes:} This table reports the coefficients corresponding to Figure~\ref{fig:nba}. For brevity, coefficients for crash, weather, holiday, season dummy, and day-of-week dummy indicators are omitted; this amounts to 1{,}980 omitted coefficients across the 180 regressions reported in the table. Predicted win indicates a point spread of --6.75 or less (negative spreads indicate the number of points a team is expected to win by); predicted close indicates a point spread between --6.75 and +4.75 exclusive; and predicted loss indicates a spread of +4.75 or more. The sample is limited to regular-season NBA games between the 2015/16 and 2018/19 seasons, encompassing four total seasons. The analysis includes home games played by Florida's two NBA teams---the Miami Heat and the Orlando Magic---with scheduled tip-off times of 6:00 PM or later. Regressions include TMC-code fixed effects, as well as month time fixed effects, season dummies, and day-of-week dummies to control for temporal variation. Standard errors are clustered at the TMC-code level. Average vehicle speed represents observed speed across road links within 5 km of each arena, broken up into 1-km distance bands, measured in 10-minute intervals from 60 minutes before to 120 minutes after each game.
  \end{spacing}
\end{minipage}
\endgroup

\addtocounter{table}{-1}
\clearpage

\begin{table}[htbp]
\centering
\refstepcounter{table}
\phantomsection
\label{tab:wildboot}
\captionsetup{justification=centering, font=small, labelfont=bf, labelsep=colon}
\caption*{\textbf{Table \thetable}\\[4pt]Wild cluster bootstrap p-values for key parameters at 2--3 km distance band}
\vspace{6pt}
\scriptsize
\setlength{\tabcolsep}{4pt}
\renewcommand{\arraystretch}{1}
\begin{tabular*}{\textwidth}{@{\extracolsep{\fill}}lcccccc@{}}
\toprule
& \multicolumn{2}{c}{\textbf{Predicted close $\times$ lose}} & \multicolumn{2}{c}{\textbf{Predicted lose $\times$ win}} & \multicolumn{2}{c}{\textbf{Predicted win $\times$ lose}} \\
\cmidrule(lr){2-3} \cmidrule(lr){4-5} \cmidrule(lr){6-7}
\textbf{Post-game window} & \textit{p} (entity) & \textit{p} (boot) & \textit{p} (entity) & \textit{p} (boot) & \textit{p} (entity) & \textit{p} (boot) \\
\midrule
+10 min & 0.029 & 0.246 & 0.013 & 0.005 & 0.000 & 0.000 \\
+20 min & 0.013 & 0.100 & 0.057 & 0.006 & 0.000 & 0.000 \\
+30 min & 0.195 & 0.135 & 0.876 & 0.000 & 0.000 & 0.000 \\
+40 min & 0.083 & 0.073 & 0.004 & 0.000 & 0.000 & 0.000 \\
+50 min & 0.034 & 0.022 & 0.002 & 0.000 & 0.000 & 0.000 \\
+60 min & 0.010 & 0.010 & 0.001 & 0.000 & 0.000 & 0.000 \\
+70 min & 0.000 & 0.000 & 0.012 & 0.000 & 0.000 & 0.000 \\
+80 min & 0.000 & 0.000 & 0.004 & 0.000 & 0.000 & 0.000 \\
+90 min & 0.000 & 0.000 & 0.082 & 0.000 & 0.000 & 0.000 \\
+100 min & 0.000 & 0.006 & 0.141 & 0.000 & 0.000 & 0.000 \\
+110 min & 0.000 & 0.068 & 0.623 & 0.000 & 0.000 & 0.000 \\
+120 min & 0.000 & 0.340 & 0.192 & 0.000 & 0.000 & 0.000 \\
\bottomrule
\end{tabular*}
\noindent
\begin{minipage}{\textwidth}
  \vspace{10pt}
  \begin{spacing}{0.9}
  \scriptsize
  \noindent\textit{Notes:} The table reports entity-clustered p-values and wild cluster bootstrap p-values for the three key outcome interaction parameters at the 2--3 km distance band across all post-game time windows. Bootstrap p-values are based on 999 replications using Rademacher weights, bootstrap type `11' (CRV1), with the null imposed, clustered at the TMC-code level. The main model specification is used throughout: entity fixed effects (TMC-code), week-of-season time fixed effects, season dummies, and entity-clustered standard errors. The sample is limited to Sundays during the regular-season home games for Florida's three NFL teams across the 2015--2019 seasons, excluding the 2018 and 2019 Miami Dolphins seasons, resulting in 84 Sunday home games. With crash dummy included.
  \end{spacing}
\end{minipage}
\end{table}

\end{document}